\documentclass[final,1p,times]{elsarticle}
\usepackage{amssymb,amsmath}
\usepackage{subfig,graphicx,multicol, color}
\usepackage[usenames,dvipsnames]{xcolor}
\usepackage{balance}
\usepackage{tikz}
\usepackage{subfig}
\usepackage{cases}
\usepackage[numbers]{natbib}
\usepackage{setspace}

\newcommand{\ra}[1]{\renewcommand{\arraystretch}{#1}}
\newcommand{\myqed}{\hfill \ensuremath{\blacksquare}}

\usepackage{amsmath, amsthm, amssymb, amsfonts}
\usepackage{subfig}
\usepackage{color, colortbl}
\usepackage{hyperref}
\usepackage{algorithm, algorithmic}
\usepackage{indentfirst}
\usepackage{booktabs}
\usepackage{setspace}
\usepackage{bm}
\usepackage{amsthm, amsmath, amssymb, dsfont, epsfig, pifont, algorithmic, algorithm, psfrag, color}
\usepackage{graphicx}
\usepackage{booktabs}
\usepackage{setspace}
\usepackage{bm}
\usepackage{shortcuts_Jor2}
\usepackage{amsthm, amsmath, amssymb, dsfont, epsfig, pifont, algorithmic, algorithm, psfrag, color}
\usepackage{graphicx}
\usepackage{amssymb}
\usepackage{amsthm}
 
\usepackage{url}
   
\newcommand{\sinc}{{\rm sinc}}

\usepackage{amsmath, amsthm, amssymb, amsfonts}
\usepackage{subfig}
\usepackage{color, colortbl}
\usepackage{hyperref}
\usepackage{algorithm, algorithmic}
\usepackage{indentfirst}
\usepackage{booktabs}
\usepackage{setspace}
\usepackage{bm}
\usepackage{amsthm, amsmath, amssymb, dsfont, epsfig, pifont, algorithmic, algorithm, psfrag, color}
\usepackage{graphicx}
\usepackage{booktabs}
\usepackage{enumerate}
\usepackage{setspace}
\usepackage{bm}
\usepackage{amsthm, amsmath, amssymb, dsfont, epsfig, pifont, algorithmic, algorithm, psfrag, color}
\usepackage{graphicx}
 
\usepackage{amsthm, amsmath, amssymb, dsfont, epsfig, pifont, algorithmic, algorithm, psfrag, color}
\usepackage{graphicx}

\usepackage{array}
\usepackage{multirow}
\newcommand{\head}[1]{\textnormal{\textbf{#1}}}
\newcommand{\normal}[1]{\multicolumn{1}{l}{#1}}

\usetikzlibrary{arrows}
\tikzstyle{c} = [circle,draw=black,minimum size=5ex]
\tikzstyle{r} = [rectangle,draw=black,minimum size=5ex]
\tikzstyle{a} = [->,>=stealth']

\def\reg{{\rm\ooalign{\hfil
     \raise.07ex\hbox{\scriptsize R}\hfil\crcr\mathhexbox20D}}}

\usepackage{mathtools}

\definecolor{blue}{RGB}{0,0,0}
\definecolor{purple}{RGB}{1,0,1}
\journal{Signal Processing}
\begin{document}
\begin{frontmatter}
\title{Ad~Hoc Microphone Array Calibration: Euclidean Distance\\ Matrix Completion Algorithm and Theoretical Guarantees}

\author[1,2]{Mohammad J. Taghizadeh}
\author[2]{Reza Parhizkar}
\author[1]{Philip N. Garner}
\author[1,2]{Herv\'e Bourlard}
\author[1]{\\Afsaneh Asaei}
\address[1]{Idiap Research Institute, Martigny, Switzerland}
\address[2]{\'Ecole Polytechnique F\'ed\'erale de Lausanne (EPFL), Switzerland}
\address{Emails: \{mohammad.taghizadeh, phil.garner, herve.bourlard, afsaneh.asaei\}@idiap.ch, reza.parhizkar@epfl.ch}
%%%%%%%%%%%%%%%%%%%%%%%%%%%%%%%%%%%%%%%%%%%%%%%%%%%%%%%%%%%%%%%%%%%%%%%%%%%%%%%%%%%%%%%%%%%%%%%%%%%%%%%%%%%

\begin{abstract}
This paper addresses the problem of ad~hoc microphone array calibration where only partial information about the distances between microphones is available. We construct a matrix consisting of the pairwise distances and propose to estimate the missing entries based on a novel Euclidean distance matrix completion algorithm by alternative low-rank matrix completion and projection onto the Euclidean distance space. This approach confines the recovered matrix to the EDM cone at each iteration of the matrix completion algorithm. The theoretical guarantees of the calibration performance are obtained considering the random and locally structured missing entries as well as the measurement noise on the known distances. {\color{blue}This study elucidates the links between the calibration error and the number of microphones along with the noise level and the ratio of missing distances. Thorough experiments on real data recordings and simulated setups are conducted to demonstrate these theoretical insights.} A significant improvement is achieved by the proposed Euclidean distance matrix completion algorithm over the state-of-the-art techniques for ad hoc microphone array calibration.\newline
\end{abstract}

\begin{keyword}
Ad-hoc microphone array calibration, Diffuse noise coherence model, Cadzow algorithm, EDM cone, Euclidean distance matrix completion.\newline
\end{keyword}
\end{frontmatter}

% \newpage
\section{Introduction}

Ad hoc microphone arrays consist of a set of sensor nodes spatially distributed over the acoustic field, in an ad~hoc fashion. 
% While the distributed acquisition provides a flexible infrastructure for high quality sound acquisition, and thus require to
% % be calibrated to function in synergy. 
Processing of the data acquired with distributed sensors involves challenges attributed to the issues such as asynchronous sampling and unknown microphone positions. In this paper, {\color{blue}we assume that the recordings are synchronized} and address the problem of finding the microphone positions; this problem is referred to as \emph{microphone calibration}. The precise knowledge of the microphones positions is required for a plethora of multi-channel audio processing applications such as distant speech recognition~\cite{dsr-mag12,do2012,asaei-thesis13,DSR-book09,mohammad_broad} and source localization and separation~\cite{Brandstein2001Microphone,taghiza11,Do-thesis, asaei11,asaei_mike}. % maliou2005,  Brandstein, 

{\color{blue}Previous studies often consider activation of a (known) source signal in a specific configuration to estimate the distances between the source and microphones. The pairwise distances are then used to reconstruct the array geometry. This approach is referred to as self-calibration.} 

{\color{purple} It may be noted that the knowledge of the source signal simplifies the estimation problems. If the signal is known beforehand, the time of arrival (ToA) of the source signal for each individual microphone is obtained through cross-correlation with the given signal. Hence, the negative effects of noise and reverberation are reduced as only one of the signals is noisy. On the other hand, if the source signal is unknown, the time difference of arrival (TDoA) for a pair of microphones is estimated through cross-correlation of the two microphone signals. Although TDoA-based methods can alleviate the need for activating a specific source signal or prior knowledge on the original signal, they may be more sensitive to noise and reverberation.}

Sachar et al.~\cite{sachar05} presented a set-up using a pulsed acoustic excitation generated by five domed tweeters for measuring the transmit times and distances between speakers and microphones. Raykar et al.~\cite{raykar2005} exploited a maximum length sequence or chirp signal in a distributed computing platform. The time difference of arrival of microphone signals were computed by cross-correlation and used for estimating the microphone locations. Since the original signal is known, these techniques are robust to noise and reverberation. 

Gaubitch et al.~\cite{Gaubitch13} proposed an auto-localization method exploiting the asynchronous time-of-arrival measurements obtained from spatially distributed acoustic events. He developed an iterative rank reduction algorithm to correct for the time offsets without imposing any geometrical constraint on the placement of the microphones and sound sources. These aligned time-of-arrival measurements are then used to estimate the location of source and microphones using the bilinear optimization approach proposed in~\cite{crocco2012bilinear}. This approach requires a minimum of five microphones and thirteen sound source events.

{\color{purple} 
In an alternative approach to alleviate the requirement for a specific source signal,}
% Flanagan and Bell~\cite{flanagan01} proposed a joint source-sensor localization scheme based on the Weiss-Friedlander method where the sensor location and direction of arrival of the sources are estimated alternately until the algorithm is converged. 
Chen et al.~\cite{energy_calib} formulated an energy-based method for maximum likelihood estimation of joint source-sensor positions. These methods require several active sources and the pairwise distances are used for a nonlinear optimization to extract the array geometry.

Pollefeys and Nistre proposed a method for direct joint source and microphone localization which requires matrix factorization and solving linear equations~\cite{pollefeys2008direct}. %kuang2013complete,
Along that line, Kuang et al.~\cite{kuang2013stratified} exploited the rank constraints to determine the unknown time offset for time-of-arrival measurements. The problem is then reduced to solving a system of polynomials for extracting the location of source and microphones. The estimates are further refined using a non-linear least squares optimization to find the correct position and time delays matching the measurements of source-microphones distances. It has been shown that exploiting the structure underlying the problem through rank and polynomial constraints enables direct recovery of source and microphone positions using as few as three microphones and six sources, thus achieves a minimal case for the self-calibration problem.

Recently, McCowan and Lincoln~\cite{mccowan2008} exploited properties of a diffuse noise field model for microphone calibration; this approach alleviates the need for activating several sources. A diffuse noise field is characterized by noise signals that propagate with equal probability in all directions and its coherence is defined by the $\sinc$ function of the distance of the two microphones. The distances can thus be estimated by fitting the computed noise coherence with the $\sinc$ function in the least squares sense. Once the pairwise distances are estimated, the classic multi-dimensional scaling method is used to reconstruct the microphone array geometry~\cite{MDS}. Along similar lines, {\color{blue}Hennecke et al.~\cite{HePlFiScHa09} proposed a hierarchical approach where the compact sub-arrays are calibrated using the coherence model of a diffuse sound field. A sound signal is activated and the relative positions of the distributed arrays are determined using steered response power based source localization.}

In this paper, we use the coherence model of a diffuse field for pairwise distance estimation due to its practical assumptions for distant audio applications in reverberant enclosures~\cite{taghiza11} and no requirement for activating a specific source signal~\footnote{{\color{purple}In fact, the ambient noise typical of many enclosures provides a sound field which enables microphone calibration and no additional source signal is required to be played.}}. Estimation of the pairwise distances becomes unreliable as the distances between the microphones are increased. Hence, the goal of this paper is to enable microphone calibration when some of the pairwise distances are missing. 

{\color{blue}The problem of missing data arises when the pairwise distance of only a subset of the sensors can be measured. If a source event is activated, device malfunctioning or architectural barriers (e.g. indoor calibration) may cause the signal of the emitted sounds to reach, or be acquired, by only a subset of the sensors. Furthermore, some of sensors deployed far apart may fail to capture the source energy leading to a locality constraint in distance estimation in ad~hoc microphone arrays~\cite{SRZ03}.
In this paper, as an example use case, the local pairwise distances are measured based on the diffuse sound field coherence model. However, the proposed algorithm and theoretical results are applicable for calibration of a general ad~hoc microphone array network. The approach proposed in this paper imposes no constraint on the geometrical set up.}  

To address the problem of missing distances, we rely on the characteristics of a Euclidean distance matrix. The matrix consisting of the squared pairwise distances has very low rank (explained in Section~\ref{sec:dis-mtx}). The low-rank property has been investigated in the past years to devise efficient optimization schemes for matrix completion, i.e. recovering a low-rank matrix from randomly known entries. Cand{\`e}s et al.~\cite{candes2010} showed that a small random fraction of the entries are sufficient to reconstruct a low-rank matrix \textit{exactly.} Keshavan et al.~proposed a matrix completion algorithm known as \opt and showed its optimality~\cite{KOImpl09}. Furthermore, they proved that their algorithm is robust against noise \cite{Keshavan10-2}. Drineas et al.~\cite{Drineas} exploited the low rank property to recover the distance matrix. However, they assume a nonzero probability of obtaining accurate distances for any pair of sensors regardless of their distance. This assumption severely restricts the applicability of their result for the microphone array calibration problem.
%KMO09noise %Recht11, KMO09exact, 

In the present study, we first estimate the pairwise distances of the microphones in close proximity using the coherence model of the signals of the two microphones in a diffuse noise field using the improved method described in~\cite{taghiza13}; this approach implies a local connectivity constraint as the pairwise distances of the further microphones can not be estimated. 
We construct a matrix of all the pairwise distances with missing entries corresponding to the unknown distances. We exploit the low-rank property of the square of this matrix to enable estimation of all the pairwise distances using matrix completion approach. 

The goal of this paper is to show that exploiting the combination of the rank condition of Euclidean distance matrices (EDMs), similarity in the measured distances, and projection on the EDM cone enables us to estimate the microphone array geometry accurately from only partial measurements of the pairwise distances. To this end, we show that matrix completion is capable of finding the missing entries in our scenario and provide theoretical guarantees to bound the error for ad~hoc microphone calibration considering the local connectivity of the noisy known entries. To increase the accuracy, we incorporate the properties of EDMs in the matrix completion algorithm. We show that imposing EDM characteristics on matrix completion improves the robustness and accuracy of extraction of the ad~hoc microphone geometry.
  
The rest of the paper is organized as follows. In Section~\ref{sec:setting}, we explain how pairwise distances of the microphones are estimated using the coherence model of the diffuse noise field as an example use case of the proposed method. Section~\ref{sec:model} describes the mathematical basis and the model used for the calibration problem. The proposed Euclidean distance matrix completion algorithm is elaborated in Section~\ref{sec:proposed_alg}. Section~\ref{sec:main} is dedicated to the theoretical guarantees for ad~hoc microphone array calibration based on matrix completion. The related methods are investigated in Section~\ref{sec:topology} and the experimental analysis are presented in section~\ref{sec:experim}. Finally, the conclusions are drawn in Section~\ref{sec:conclusion}.
        
%%%%%%%%%%%%%%%%%%%%%%%%%%%%%%%%%%%%%%%%%%%%%
% RESERVE
%%%%%%%%%%%%%%%%%%%%%%%%%%%%%%%%%%%%%%%%%%%%%

% Microphone arrays are widely used in distant audio technolo-
% gies to enable source localization and separation [1–7], videocon-
% ferencing [8] and distant speech recognition in multiparty environ-
% ments [9–12]. Ad hoc arrays provide a distributed and flexible in-
% frastructure for high quality sound acquisition, and thus require to
% be calibrated to function in synergy. The focus of this paper is on the
% microphone array position calibration. This task is often achieved in
% two steps: estimation of the distances between the pairs of micro-
% phones and reconstruction of the array geometry from the pairwise
% distance information.

% Recent advances in mobile computing and communication technologies enable using cell phones, PDA's or tablets as a flexible acquisition set-up providing an ad~hoc network of microphones. However, the unknown prior information on relative positions of the microphones is a key problem to achieve effective data processing. In the following, we review some of the prior approaches for microphone array calibration. 

% In the context of unsupervised calibration where no source activation is required, the underlying approaches are largely confined to the very compact array configurations. Hence, a key contribution of the proposed approach is to address the problem of missing pairwise distances in ad~hoc array calibration.

%%%%%%%%%%%%%%%%%%%%%%%%%%%%%%%%%%%%%%%%%%%%%%%%%%%%%%%%%%%%%%%%%%%%%%%%%%%%%%%%%%%%%%%%%%%%
\section{Example Use Case}
\label{sec:setting}
%%%%%%%%%%%%%%%%%%%%%%%%%%%%%%%%%%%%%%%%%%%%%%%%%%%%%%%%%%%%%%%%%%%%%%%%%%%%%%%%%%%%%%%%%%%%
We consider $N$ microphones located at random positions on a large circular table in a meeting room with homogeneous reverberant acoustics. In the time intervals that there is no active speaker, diffuse noise is the dominant signal in the room. The table is located at the center of the room, hence deviation from diffuseness near the walls can be neglected. Based on the theory of the diffuse noise model, the distance of each two close microphones can be estimated by computing the coherence of their signals $\Gamma$, and fitting a $\sinc$ function with the relation expressed as 
\begin{equation}
\label{eq:coherence}
 \Gamma_{ij}(\omega)= \sinc\,{\left(\frac{\omega d_{ij}}{c} \right)}\;,
\end{equation}
where $\omega$ is the frequency, $d_{ij}$ is the distance between the two microphones $i$ and $j$, and $c$ is the speed of sound \cite{cook55}. Figure~\ref{fig:sinc2} represents an example of the coherence and the fitted $\sinc$ function.

% \footnote{\color{blue} The close microphones are indicated by the limitation of the diffuse noise coherence model for microphone array calibration. This problem is studied in depth in~\cite{Taghizadeh2014242} which elucidates the relation between the maximum distance that can be measured and the size of the enclosure and the reverberation level. For our experiments on real data recordings presented in this paper, the microphones are considered to be close if their distance is less than 73 cm.}

% %%%%%%%%%%  FIGURE 1   %%%%%%%%%%
% \begin{figure}[h]
% \begin{center}
% \includegraphics[width=\linewidth,height=.4\linewidth]{synt_mic8_4_196-crop}
% \end{center}
% \caption{{Coherence of the signal of two microphones at $d_{ij}=20\;$cm and the fitted $\sinc$ function using real data recordings.}}
% \label{fig:sinc2}
% \end{figure}
% %%%%%%%%%%%%%%%%%%%%%%%%%%%%%%%%%

In practice, if the distance between the sensors is large (e.g. greater than $73\,$cm~\cite{taghiza13,Taghizadeh2014242}) we observe deviations from the diffuse characteristics. The maximum distance that can be computed by this method is assumed to be $d_{max}$. Therefore, pairwise distances greater than $d_{max}$ are missing implying a locality structure in the missing entries in the distance matrix consisting of the pairwise distances. In addition, the computation algorithm can lead to deviation from the model resulting in unreliable estimates of the short distances causing random missing entries in the distance matrix; {\color{blue} the random missing entries intend to model the distances which can not be measured due to mismatch or violations of the underlying pairwise distance estimation model pertained to the acoustic ambiguities. Furthermore, the known entries are noisy due to measurement inaccuracies and variations of diffuseness~\cite{Taghizadeh2014242}.}

%%%%%%%%%%%%%%%%%%%%%%%%%%%%%%%%%%%%%%%%%%%%%%%%%%%%%%%%%%%%%%%%%%%%%%%%%%%%%%%%%%%%%%%%%%%%
\section{Problem Formulation}
\label{sec:model}
%%%%%%%%%%%%%%%%%%%%%%%%%%%%%%%%%%%%%%%%%%%%%%%%%%%%%%%%%%%%%%%%%%%%%%%%%%%%%%%%%%%%%%%%%%%%
%%%%%%%%%%%%%%%%%%%%%%%%%%%%%%%%%%%%%%%%%%%%%%%%%%%%%%%%%%%%%%%%%%%%%%%%%%%%%%%%%%%%%%%%%%%%
\subsection{Distance Matrix}
\label{sec:dis-mtx}
%%%%%%%%%%%%%%%%%%%%%%%%%%%%%%%%%%%%%%%%%%%%%%%%%%%%%%%%%%%%%%%%%%%%%%%%%%%%%%%%%%%%%%%%%%%%
Consider a distance matrix $\D_{N \times N}$ consisting of the distances between $N$ microphones constructed as
\begin{equation}
\D = \left[d_{ij}\right] , \quad d_{ij}=\norm{\bm{x}_i-\bm{x}_j}, \quad  i,j \in \{1,\ldots, N\}\;,
\label{eq:dist_mat}
\end{equation}
where $d_{ij}$ is the Euclidean distance between microphones $i$ and $j$ located at $\bm{x}_i$ and $\bm{x}_j$. Therefore, $\D$ is a symmetric matrix and it is often full rank.

Let $\X_{N\times {\zeta}}$ denote the position matrix whose $i^{\text{th}}$ row, $\mathbi{x}_i^T \in \R^\zeta$, is the position of microphone $i$ in $\zeta$-dimensional Euclidean coordinate where microphones are deployed and $.^T$ denotes the transpose operator. By squaring the elements of $\D$, we construct a matrix $\M_{N \times N}$ which can be written as
\begin{equation}\label{eq:rankd4}
\M = \ones_N \bm{\Lambda}^T + \bm{\Lambda} {\ones_N}^T -2\X \X^T\;,
\end{equation}
where $\ones_N \in \R^N$ is the all ones vector and $\bm{\Lambda} = (\X \circ \X)\ones_\zeta$; $\circ$ denotes the Hadamard product. We observe that $\M$ is the sum of three matrices of rank 1, 1 and at most $\zeta$ respectively. Therefore, the rank of the squared distance matrix constructed of the elements $\M_{ij} = \left[d_{ij}^2\right]$ is at most $\zeta+2$~\cite{Drineas}. For instance, if the microphones are located on a plane or shell of a sphere, $\M$ has rank 4 and if they are placed on a line or circle, the rank is exactly 3. Hence, there is significant dependency between the elements of $\M$ and exploiting this low-rank property is the core of the proposed algorithm in this paper.

%%%%%%%%%%%%%%%%%%%%%%%%%%%%%%%%%%%%%%%%%%%%%%%%%%%%%%%%%%%%%%%%%%%%%%%%%%%%%%%%%%%%%%%%%%%%
\subsection{Objective}
\label{sec:pro-state}
%%%%%%%%%%%%%%%%%%%%%%%%%%%%%%%%%%%%%%%%%%%%%%%%%%%%%%%%%%%%%%%%%%%%%%%%%%%%%%%%%%%%%%%%%%%%
The noisy estimates of the pairwise distances are modeled as 
\begin{equation}\label{eq:w-noise}
 \td_{ij} = d_{ij}+w_{ij} \quad ; \quad \tD = \D + \W \;,
\end{equation}
where $w_{ij}$ is the measurement noise for distance $d_{ij}$ and $\W$ is the corresponding measurement noise matrix. We introduce a noise matrix on the squared distance matrix as 
\begin{equation}\label{eq:def_Z}
\Z = \tM - \M = \tD \circ \tD- \D \circ \D \;,
\end{equation}
where $\tM$ is the noisy squared distance matrix.
 
As described in Section~\ref{sec:setting}, there are two kinds of missing entries. The first group consists of the structured missing entries corresponding to the distances greater than $d_{max}$. We denote this group by $S$ defined as
\begin{equation}\label{eq:def_S}
S = \{(i,j):\; d_{ij} \geq d_{max} \}\;,
\end{equation}
These structured missing entries are represented by a matrix
\begin{equation}\label{eq:def_Dbs}
	\Ds_{ij} = 
		\begin{cases}
			\D_{ij} & \text{if } (i,j) \in S\\
			0 & \text{otherwise}
		\end{cases}
\end{equation}
Hence, the noiseless recognized pairwise distance matrix is given by  
\begin{equation}
\label{eq:Dbsbar}
	\Dsb=\D-\Ds\;,
\end{equation}
and we obtain the corresponding known squared distance matrix as
\begin{equation}\label{eq:Mbsbar1}
\begin{split}
& \Ms = \Ds \circ \Ds \\
& \Msb = \Dsb \circ \Dsb = \M-\Ms\;.
\end{split}
\end{equation}
Considering the noise on the known entries, we obtain 
\begin{equation}\label{eq:Mbsbar2}
	\tMsb= \Msb + \Zsb\;,
\end{equation}
where $\Zsb$ denotes the noise on the known entries in the squared distance matrix.
 
To model the random missing entries, we assume that each entry is sampled with probability $p$; sampling can be introduced by a projection operator on an arbitrary matrix $\Q_{N \times N}$, given by 
\begin{equation}\label{eq:def_Pc}
\Pc_E(\Q)_{ij} = 
	\begin{cases}
		\Q_{ij} & \text{if } (i,j) \in E\\
		0 & \text{otherwise}
	\end{cases}
\end{equation}
where $E\subseteq [N]\times[N]$ denotes the known entries after random erasing process and has cardinality $|E| \approx pN^2$. Therefore, the final known squared distance matrix is given by
\begin{equation}\label{eq:MhatE}
\M^E= \Pc_E(\tMsb)\;.
\end{equation}
The goal of the matrix recovery algorithm is to find the missing entries and remove the noise, given matrix $\M^E$.

%%%%%%%%%%%%%%%%%%%%%%%%%%%%%%%%%%%%%%%%%%%%%%%%%%%%%%%%%%%%%%%%%%%%%%%%%%%%%%%%%%%%%%%%%%%%
\subsection{Noise Model}
\label{sec:noise}
%%%%%%%%%%%%%%%%%%%%%%%%%%%%%%%%%%%%%%%%%%%%%%%%%%%%%%%%%%%%%%%%%%%%%%%%%%%%%%%%%%%%%%%%%%%%
The level of noise in extracting the pairwise distances, $w_{ij}$ in~\eqref{eq:w-noise}, increases as the distances become larger~\cite{Taghizadeh2014242}. We model this effect through
\begin{equation}
\label{eq:output}
\W = {\bf\Upsilon} \circ \D\;,
\end{equation}
where the normalized noise matrix ${\bf\Upsilon}_{N \times N}$ consists of entries with sub-Gaussian distribution with variance $\varsigma^2$, thus~\cite{Keshavan10-2} 
\begin{equation}\label{eq:def_Up}
 \prob(|{\bf\Upsilon}_{ij}| \geq \beta) \leq 2\, \text{e}^{-\frac{\beta^2}{2 \varsigma^2}}\;.
\end{equation}
Based on~\eqref{eq:Mbsbar2}, $\Zsb_{ij} = 2d_{ij}^2 {\bf\Upsilon}_{ij} + d_{ij}^2 {\bf\Upsilon}_{ij}^2$; thereby $\Zsb_{ij}$ is also a sub-Gaussian random variable with a bounded constant $2\varsigma d_{ij}^2$. The physical setup confines $|\Zsb_{ij}| \leq 4a^2$ where $a$ is the radius of the table\footnote{{\color{blue}The sub-Gaussian assumption is exploited for the proof of Theorem~\ref{thm:main3} stated in Section~\ref{sec:main}. This model is not restrictive in practice and a Gaussian noise is considered for the simulations conducted in Section~\ref{sec:experim}.}}.

%%%%%%%%%%%%%%%%%%%%%%%%%%%%%%%%%%%%%%%%%%%%%%%%%%%%%%%%%%%%%%%%%%%%%%%%%%%%%%%%%%%%%%%%%%%%%%%%%
\subsection{Evaluation Measure}
\label{sec:evaluation}
%%%%%%%%%%%%%%%%%%%%%%%%%%%%%%%%%%%%%%%%%%%%%%%%%%%%%%%%%%%%%%%%%%%%%%%%%%%%%%%%%%%%%%%%%%%%%%%%%%%
Extracting the absolute position of the microphones deployed in $\zeta$ dimensional space requires at least $\zeta+1$ anchor points in addition to the distance matrix. Therefore, in a scenario where the only available information is pairwise distances, the evaluation measure must quantify the error in estimation of the \emph{relative} position of the microphones thus robust to the rigid transformations (translation, rotation and reflection). Hence, we quantify the distance between the actual locations $\X$ and estimated locations $\hX$ as~\cite{book-MDS}
\begin{equation}\label{distance}
\begin{split}
\text{dist}(\X,\hX) = & \frac{1}{N}\norm{\J \X\X^T\J-\J\hX\hX^T\J }_\text{F} \;,\\
& \J=\ind_N-(1/N)\ones_N\ones_N^T
\end{split}
\end{equation}
where $\norm{\cdot}_\text{F}$ denotes the Frobenius norm and $\ind_N$ is the $N \times N$ identity matrix. The distance measure stated in \eqref{distance} is useful to compare the performance of different methods in terms of microphone array geometry estimation. \newline

Table~\ref{tab:notations} summarizes the set of important notation. % used in the paper.

% \begin{table*}[tb]
% \caption{Summary of the notation.\vspace{0.2cm}}
% \ra{0.8}
% % \centering
% \small
% \begin{tabular}{@{}ll | ll@{}}
% \toprule
% Symbol & Meaning & Symbol & Meaning\\
% \midrule
% $N$ & number of microphones & $\D$ & complete noiseless distance matrix\\
% $a$ & radius of the circular table on which microphones are distributed & $\M$ & squared distance matrix\\
% $\varsigma$ & normalized standard deviation of noise & $\tM$ & noisy squared distance matrix\\	
% $\Pc_E$ & projection into matrices with entries on index set $E$ & $\hM$ & estimated squared distance matrix\\
% $\cP_e$ & projection to EDM cone & $\Z$ & noise matrix \\
% $p$ & probability of having random missing entries  & $\M^E$ & observed matrix\\
% $d_{max}$ & radius of the circle defining structured observed entries & $\X$ &positions matrix\\
% $\Ms$ & distance matrix with observed entries on index set $S$ & $\hX$ &estimated positions matrix\\
% \bottomrule
% \end{tabular}
% \label{tab:notations}
% \end{table*}

%%%%%%%%%%%%%%%%%%%%%%%%%%%%%%%%%%%%%%%%%%%%%%%%%%%%%%%%%%%%%%%%%%%%%%%%%%%%%%%%%%%%%%%%%%%%
\section{Euclidean Distance Matrix Completion Algorithm}
\label{sec:proposed_alg}
%%%%%%%%%%%%%%%%%%%%%%%%%%%%%%%%%%%%%%%%%%%%%%%%%%%%%%%%%%%%%%%%%%%%%%%%%%%%%%%%%%%%%%%%%%%%
The approach proposed in this paper exploits low-rank matrix completion and incorporates the EDM properties for recovering the distance matrix.

%%%%%%%%%%%%%%%%%%%%%%%%%%%%%%%%%%%%%%%%%%%%%%%%%%%%%%%%%%%%%%%%%%%%%%%%%%%%%%%%%%%%%%%%%%%%
\subsection{Matrix Completion}
%%%%%%%%%%%%%%%%%%%%%%%%%%%%%%%%%%%%%%%%%%%%%%%%%%%%%%%%%%%%%%%%%%%%%%%%%%%%%%%%%%%%%%%%%%%%
We recall our problem of having $N$ microphones distributed on a space of dimension $\zeta$. Hence, the squared distance matrix $\M$ has rank $\eta=\zeta+2$, but it is only partially known. The objective is to recover $\M_{N \times N}$ of rank $\eta \ll N $ from a sampling of its entries without having to ascertain all the $N^2 $ entries, or collect $N^2$ measurements about $\M$. The approach proposed through \emph{matrix completion} relies on the fact that a low-rank data matrix carries much less information than its ambient dimension implies. Intuitively, as the matrix $\M$ has $(2N-\eta) \eta$ degrees of freedom\footnote{The degrees of freedom can be estimated by counting the parameters in the singular value decomposition (the number of degrees of freedom associated with the description of the singular values and of the left and right singular vectors). When the rank is small, this is considerably smaller than $N^2$~\cite{candes12}.}, we need to know at least $\eta N$ of the row entries as well as $\eta N$ of the column entries reduced by $\eta^2$ of the repeated values to recover the entire elements of $\M$. 

Given $\M^E$ defined in~\eqref{eq:MhatE}, the matrix completion recovers an estimate of the distance matrix $\hat{\M}$ through the following optimization
\begin{equation}
\begin{split}
&\text{Minimize} \quad  \quad \quad \text{rank}\,(\hat{\M}\ ) \quad \quad \quad \quad\\
&\text{subject  to}  \quad \quad{\hat{\M} }_{ij} = \M_{ij}\,, \quad \quad (i,j) \in E
\end{split}
\end{equation}

In this paper, we use the procedure of \opt proposed by Keshavan et al. \cite{Keshavan10-2} for estimating a matrix given the desired rank $\eta$. 
This algorithm is implemented in three steps: (1) Trimming, (2) Projection and (3) Minimizing the cost function. 

In the trimming step, a row or a column is considered to be over-represented if it contains more samples than twice the average number of {\color{blue}non-zero} samples per row or column. These rows or columns can dominate the spectral characteristics of the observed matrix $\M^E$. Thus, some of their entries are removed uniformly at random from the observed matrix. Let $\tM^E$ be the resulting matrix of this trimming step. 

In the projection step, we first compute the singular value decomposition (SVD) of $\tM^E$ thus
\begin{equation}
\tM^E = \sum_{i=1}^{N}\sigma_i(\tM^E)  \bm{U}_{.i} \bm{V}_{.i}^T\;,
\end{equation}
where $\sigma_i(\cdot)$ denotes the $i^{\text{th}}$ singular value of the matrix and $\bm{U}_{.i}$ and $\bm{V}_{.i}$ designate the $i^{\text{th}}$ column of the corresponding SVD matrices. 
Then, the rank-$\eta$ projection, $\cP_\eta(\cdot)$ returns the matrix obtained by setting to $0$ all but the $\eta$ largest singular values as   
\begin{equation}
\label{q-projection}
\cP_\eta(\tM^E) = (N^2/|E|)\sum_{i=1}^{\eta}\sigma_i(\tM^E)\bm{U}_{.i} \bm{V}_{.i}^T=  \U_0\S_0\V_0^T\;.
\end{equation}
Starting from the initial guess provided by the rank-$\eta$ projection $\cP_\eta(\tM^E)$, $\U = \U_0$ , $\V = \V_0$ and $\S = \S_0$, 
the final step solves a minimization problem stated as follows: 
Given $\U \in \R^{N\times \eta}, \V \in \R^{N\times \eta}$, find
\begin{equation}\label{eq:mc1}
\begin{split}
 F(\U,\V) &= \min_{\S\in\R^{\eta\times \eta}} \mathcal{F}(\U,\V,\S)\,,\\
 \mathcal{F}(\U,\V,\S) &= \frac{1}{2}\sum_{(i,j)\in E} (\M_{ij}- (\U\S\V^T)_{i,j})^2
\end{split}
\end{equation}
$F(\U,\V)$ is determined by minimizing the quadratic function $\mathcal{F}$ over $\S$, $\U$, $\V$ estimated by gradient decent with line search in each iteration. This last step tries to get us as close as possible to the correct low-rank matrix $\M$. 

%%%%%%%%%%%%%%%%%%%%%%%%%%%%%%%%%%%%%%%%%%%%%%%%%%%%%%%%%%%%%%%%%%%%%%%%%%%%%%%%%%%%%%%%%%%%%%%%%%%%
\subsection{Cadzow Projection to the Set of EDM Properties}
\label{sec:cadzow}
%%%%%%%%%%%%%%%%%%%%%%%%%%%%%%%%%%%%%%%%%%%%%%%%%%%%%%%%%%%%%%%%%%%%%%%%%%%%%%%%%%%%%%%%%
The classic matrix completion algorithm as described above recovers a low-rank matrix with elements as close as possible to the known entries. However, the recovered matrix does not necessarily correspond to a Euclidean distance matrix; for example, EDMs are symmetric with zero diagonal elements. These properties are not incorporated in the matrix completion algorithm. Hence, we modify the aforementioned procedure to have, as output, matrices that are closer to EDMs~\cite{taghiza13}. 

To this end, we apply a Cadzow-like method. The Cadzow algorithm \cite{Cadzow88} (also known as Papoulis-Gershberg) is a method for finding a signal which satisfies a composite of properties by iteratively projecting the signal into the property sets. We modify the matrix completion algorithm by inserting an extra step at each iteration. In the classic version of this algorithm a simple rank-$\eta$ approximation is used as the starting point for the iterations using gradient descent on \eqref{eq:mc1}. After each iteration of the gradient descent, we apply the transformation $\cP_{c}: \R^{N\times N} \longmapsto \mathbb{S}_h^N$ on the obtained matrix where $\mathbb{S}_h^N$ is the space of symmetric, positive hollow matrices, to make sure that the output satisfies the following properties
\begin{equation}
\hM \in \mathbb{S}_h^N \Longleftrightarrow
\begin{cases}
 d_{ij} = 0 \Leftrightarrow {\bm x_i} = {\bm x_j}\\
 d_{ij} > 0,\; i \neq j\\
 d_{ij}=d_{ji}
\end{cases}
\end{equation}
for $i,j \in [N]$; nonnegativity and symmetry are achieved by setting all the negative elements to zero and averaging the symmetric elements.  

%%%%%%%%%%%%%%%%%%%%%%%%%%%%%%%%%%%%%%%%%%%%%%%%%%%%%%%%%%%%%%%%%%%%%%%%%%%%%%%%%%%%%%%%%%%%%%%%%%%%
\subsection{Matrix Completion with Projection onto the EDM cone}
\label{sec:EDMcone}
%%%%%%%%%%%%%%%%%%%%%%%%%%%%%%%%%%%%%%%%%%%%%%%%%%%%%%%%%%%%%%%%%%%%%%%%%%%%%%%%%%%%%%%%%
In section \ref{sec:cadzow}, three characteristics of EDMs are employed through the Cadzow projection to reduce the reconstruction error of the distance matrix. In order to increase the accuracy even further, we propose to project to the cone of Euclidean distance matrix, $\mathbb{EDM}^N$, at each iteration of the algorithm. In other words, after one step of the gradient descent method on the Cartesian product of two Grassmannian manifolds $\mathcal{G}$, we apply a projection, $\cP_{e}: \R^{N\times N} \longmapsto \mathbb{EDM}^N$ to decrease the distance between the estimated matrix and the EDM cone. This is visualized in Figure \ref{fig:grassman}. Note that the illustration of the cone and the manifold are not mathematically accurate and only serve as visualizations (The dimension of the cone and the manifold are too large to be illustrated graphically).

% %%%%%%%%%%%   FIGURE  2   %%%%%%%%%%%%%%%%%%%%%%%%%
% \begin{figure}[tb]
% \centering
% \includegraphics[width=0.6\linewidth]{cone_etc}
% \caption{Matrix completion with projection onto the EDM cone. }
% \label{fig:grassman}
% \end{figure}
% %%%%%%%%%%%%%%%%%%%%%%%%%%%%%%%%%%%%%%%%%%%%%%%%%%%
The projected matrix must satisfy the following EDM properties~\cite{edm-prop} 
\begin{equation}\label{eq:cone_edm}
	\hM \in \mathbb{EDM}^N \Longleftrightarrow 
		\begin{cases}
		      -z^T \hM z \geq 0 \\
		      \ones^T z = 0 \\
	              (\forall \|z\|=1)\\
		      \hM \in \mathbb{S}_h^N
		\end{cases}
\end{equation}

The EDM properties include the triangle inequality, thus 
\begin{equation}
 d_{ij} \leq d_{ik} + d_{kj}, \quad i \neq j \neq k\;,
\end{equation}
as well as the relative-angle inequality; $\forall i, j, l \neq k \in [N], i < j < l$, and for $N \geq 4$ distinct points $\{{\bm x_k} \}$, the inequalities  
\begin{equation}
\begin{split}
 \text{cos} (\tau_{jkl} + &\tau_{lkj}) \leq \text{cos}\, \tau_{ikj} \leq \text{cos} (\tau_{ikl} - \tau_{lkj}) \\
 & 0 \leq \tau_{ikl},\, \tau_{lkj},\, \tau_{ikj} \leq \pi
\end{split} 
\end{equation}
where $\tau_{ikj}$ denotes the angle between vectors at ${\bm x_k}$ and it is satisfied at each position ${\bm x_k}$. 

The projection $\cP_{e}$ must map the output of matrix completion to the closest matrix on $\mathbb{EDM}^N$ with the properties listed in~\eqref{eq:cone_edm}. The projection onto $\mathbb{S}_h^N$ is achieved by $\cP_{c}$ implemented via Cadzow; thereby, we define $(\U_c, \V_c, \S_c) =  \cP_{c} (\U^{k+1/2}, \V^{k+1/2}, \S^{k+1/2})$. To achieve the full EDM properties, we search in the EDM cone using a cost function defined as
\begin{equation}\label{eq:edm-cost}
 \mathcal{H}(\X) = \left\| \ones_N \bm{{\Lambda}}^T  + \bm{{\Lambda}} {\ones_N}^T -2\X \X^T - \U_c \S_c \V_c^T \right\|^2_\text{F}\;.
\end{equation}

To minimize the cost function, we start from the vertex of the $\mathbb{EDM}^N$ thus assume that all microphones are located in the origin of the space $\R^\zeta$. Denoting the location of microphone $i$ with $\bm{x}_i = [x_{i1},...,x_{i\zeta}]^T$, $\mathcal{H}(\X)$ is a polynomial function of $x_{i1}$ of degree $4$. The minimum of $\mathcal{H}(\X)$ with respect to $x_{i1}$ can be computed by equating the partial derivation of equation~\eqref{eq:edm-cost} to zero to obtain the new estimates, thus
\begin{equation}
\begin{split}
  \hX & = \argmin_{\X} \mathcal{H}(\X) \\
 (\U^{k+1}, \V^{k+1}, \S^{k+1}) & =  \text{SVD}\; (\ones_N \bm{\hat{\Lambda}}^T  + \bm{\hat{\Lambda}} {\ones_N}^T -2\hX \hX^T)
\end{split}
\end{equation}
where $\bm{\hat{\Lambda}} = (\hX \circ \hX)\ones_\zeta$. The stopping criteria is satisfied when the new estimates differ from the old ones by less than a threshold. 

The modified iterations can be summarized in two steps:
\begin{itemize}%[leftmargin=0cm,itemindent=.5cm]
\item iteration $k+1/2$: 
\begin{equation}
%\footnotesize{
\begin{split}
&\U^{k+1/2}= \U^k + \vartheta \frac{\partial F(\U^k,\V^k)}{\partial \U}\\
&\V^{k+1/2}= \V^k + \vartheta \frac{\partial F(\U^k, \V^k)}{\partial \V}\\
&\S^{k+1/2}= \argmin_{\S} \mathcal{F}(\U^k, \V^k, \S)\,
%  &(\U^{k+1/2}, \V^{k+1/2}, \S^{k+1/2}) = \\ 
%  & (\U^k + \vartheta \frac{\partial F(\U^k,\V^k)}{\partial \U}, \V^k + \vartheta \frac{\partial F(\U^k, \V^k)}{\partial \V}, \argmin_{\S} \mathcal{F}(\U^k, \V^k, \S))\,.
\end{split} %}
\end{equation}
\item iteration $k+1$:
\begin{equation}\label{eq:def-pe-proj}
(\U^{k+1}, \V^{k+1}, \S^{k+1}) =  \cP_{e}  (\U^{k+1/2}, \V^{k+1/2}, \S^{k+1/2})
\end{equation}
\end{itemize}
where $\vartheta$ is the step-size found using line search. 

Once the distance matrix is recovered by either classic or Cadzow matrix completion algorithms, MDS is used to find the coordinates of the microphones, $\hX$, whereas the proposed Euclidean distance matrix completion algorithm directly yields the coordinates. 

%%%%%%%%%%%%%%%%%%%%%%%%%%%%%%%%%%%%%%%%%%%%%%%%%%%%%%%%%%%%%%%%%%%%%%%%%%%%%%%%%%%%%%%%%%%%%%%%%%%%
\section{Theoretical Guarantees for Microphone Calibration}
\label{sec:main}
%%%%%%%%%%%%%%%%%%%%%%%%%%%%%%%%%%%%%%%%%%%%%%%%%%%%%%%%%%%%%%%%%%%%%%%%%%%%%%%%%%%%%%%%%%%%%%%%%%%%
In this section, we derive the error bounds on the reconstruction of the positions of $N$ microphones distributed randomly on a circular table of radius $a$ using the matrix completion algorithm and considering the locality constraint on the known entries, i.e. $d_{ij} \leq d_{max}$, as well as the noise model with the standard deviation $\varsigma\, d_{ij}$ as stated in~\eqref{eq:def_Up}. Based on the following theorem we guarantee that there is an upper bound on the calibration error which decreases by the number of microphones. 
\begin{thm}\label{thm:main1}
There exist constants $C_1$ and $C_2$, such that the output $\hX$ satisfies 
\begin{equation}
\label{eq:mainthm1}
\text{dist}(\X,\hX)\leq C_1 \frac{a^2\log_2N}{pN} + C_2 \varsigma \frac{d^2_{max}}{\sqrt{pN}}
\end{equation}
with probability greater than $1 - N^{-3}$, provided that the right-hand side is less than $\sigma_\eta(\M)/N$.
\end{thm}

%%%%%%%%%%%%%%%%%%%%%%%%%%%%%%%%%%%%%%%%%%%%%%%%%%%%%%%%%%%%%%%%%%%%%%%%%%%%%%%%%%%%%%%%%%%%%%%%%
\subsection{Proof of Theorem 1}
\label{sec:proof1}
%%%%%%%%%%%%%%%%%%%%%%%%%%%%%%%%%%%%%%%%%%%%%%%%%%%%%%%%%%%%%%%%%%%%%%%%%%%%%%%%%%%%%%%%%%%%%%%%%%%
% The important specification in using the aforementioned  matrix completion method is the \textit{incoherence} property in matrix $\M$.
The squared distance matrix $\M \in \R^{N \times N} $ with rank$-\eta$, singular values $\sigma_k(\M),\, k\in [\eta] $ and singular value decomposition $\U \Si \U^T$ is \textit{$(\mu_1, \mu_2)$-incoherent} if the following conditions hold.
\begin{itemize} 
\item[{$\mathcal{A}_1$.}] For all  $i\in [N]$: $\sum_{k=1}^{\eta}{\U_{ik}^2} \le \eta\, \mu_1 $\;.
\item[{$\mathcal{A}_2$.}] For all $i,j\in[N]$: $\big| \sum_{k=1}^{\eta}{\U_{ik}(\sigma_k(\M)/\sigma_1(\M))\U_{jk}} \big|\leq\sqrt{\eta} \, \mu_2$\;.
\end{itemize}
where without loss of generality, $\U^T \U = N \I$. 

For a $(\mu_1, \mu_2)$-incoherent matrix $\M$,~\eqref{eq:main} is correct with probability greater than $1-N^{-3}$; cf.~\cite{Keshavan10-2}-Theorem~1.2. 
\begin{equation}
\label{eq:main}
\frac{1}{N} \| \M - \hM \|_\text{F} \leq \frac{C^{\prime}_1\,\|\Pc_E(\M^s)\|_2 + C^{\prime}_2\,\norm{\Pc_E(\Zsb)}_2}{p\,N}\;,
\end{equation}
provided that
\begin{equation}\label{eq:if1}
% \small
 |E| \geq C^{\prime}_1 N \kappa_\eta^2(\M) \; \text{max} \left\{ \mu_1 \eta \log N \, ; \, \mu_1^2 \eta^2 \kappa_\eta^4(\M) \, ; \, \mu_2^2 \eta^2 \kappa_\eta^4(\M) \right\}\;,	
\end{equation}
and  
\begin{equation}\label{eq:if2}
% \small
\frac{C^{\prime}_1\,\|\Pc_E(\M^s)\|_2 + C^{\prime}_2\,\norm{\Pc_E(\Zsb)}_2}{p\,N} \leq \sigma_\eta(\M)/N\;,
\end{equation}
where the condition number $\kappa_\eta(\M) = \sigma_1(\M)/\sigma_\eta(\M)$.

To prove Theorem \ref{thm:main1}, in the first step, we show the correctness of the upper bound stated in \eqref{eq:mainthm1} based on the following Theorems \ref{thm:main2} and \ref{thm:main3}. In the second step, conditions \eqref{eq:if1} and \eqref{eq:if2} are shown to hold along with the $(\mu_1,\mu_2)$-incoherence property.
\begin{thm}\label{thm:main2}
There exists a constant $C^{\prime \prime}_1$, such that with probability greater than $1 - N^{-3}$,
\begin{equation}
\|\Psi_E(\M^s)\|_2 \leq C^{\prime \prime}_1\, a^2\, \log_2N\;.
%\|\cP_E(\bD^s)\|_2 \leq \frac{C \delta^3 \,r^2 \,p\,(\log n)^3}{n^2} \,,
\end{equation}
The proof of this theorem is explained in Appendix 1. %~\ref{proof:thm2-app}.
% with probability greater than $1 - N^{-3}$. %, where $\M^s$ and $\Psi_E(\cdot)$ are defined in \eqref{eq:Mbsbar1} and \eqref{eq:def_Pc}.
\end{thm}

\begin{thm}
\label{thm:main3}
There exists a constant $C^{\prime \prime}_2$, such that with probability greater than $1 - N^{-3}$,
\begin{equation}
\label{eq:noiseb}
 \norm{\Pc_E(\Zsb)} \leq C^{\prime \prime}_2 d_{\max}^2 \varsigma \sqrt{pN}\;.
\end{equation}
The proof of this theorem is explained in Appendix 2. %\ref{proof:thm3-app}. 
\end{thm}

On the other hand, the following condition holds for any arbitrary network of microphones~\cite{reza-thesis}  %reza13,  
\begin{equation}\label{eq:oldthm1}
\text{dist}(\X,\hX)\leq \frac{1}{N}|| \M - \hM ||_\text{F}\;.
\end{equation}
Therefore, based on Theorem~\ref{thm:main2}, Theorem~\ref{thm:main3} and the relations \eqref{eq:main} and~\eqref{eq:oldthm1}, the upper bound stated in \eqref{eq:mainthm1} is correct where $C_1 = C^{\prime}_1 C^{\prime \prime}_1$ and $C_2 = C^{\prime}_2 C^{\prime \prime}_2$; it is enough to investigate conditions \eqref{eq:if1} and \eqref{eq:if2} and $(\mu_1,\mu_2)$-incoherency of $\M$ to prove Theorem \ref{thm:main1}.

To show the inequality stated in \eqref{eq:if1}, we can equivalently show that 
\begin{equation}\label{eq:if11}
 Np\geq C^\prime_1\mu^2 \eta^2 \kappa_\eta^6(\M)\log N	\;,
\end{equation}
where $\mu = \max (\mu_1,\mu_2)$. In order to show that \eqref{eq:if11} holds with high probability for $N \geq \mathcal{C} \log N/p$ and some constant $\mathcal{C}$, we show that $\kappa_\eta(\M)$ and $\mu$ are bounded with high probability independent of $N$. 

The squared distance between $\bm{x}_i$ and $\bm{x}_j \in \R^\zeta$ is given by
\begin{equation}
 \M_{ij}= \rho_i^2+\rho_j^2-2\mathbi{x}_i^T \mathbi{x}_j \;,
\end{equation}
where $\rho_i$ is the distance of microphone $i$ from the center of the table. The squared distance matrix can be expressed as   
\begin{equation}\label{eq:asa}
\M = \A {\bf \mathcal{S}} \A^T \;,
\end{equation}
where for a planar deployment of microphones, i.e., $\zeta=2,\; \eta=4$, and $\mathbi{x}_i^T = [x_i, y_i] \in \R^2$, we have
\begin{equation}
\A = \begin{bmatrix}a/2&x_{1}&y_{1}&-a^2/4+\rho_1^2\\ \vdots&\vdots&\vdots & \vdots\\ a/2&x_{N}&y_{N} &-a^2/4+\rho_N^2 \end{bmatrix} \;,
\end{equation}
and
\begin{equation}
{\bf \mathcal{S}} = \begin{bmatrix}2&0&0&2/a\\0&-2&0&0\\0&0&-2&0\\2/a&0&0&0\end{bmatrix}\;.
\end{equation}
Since ${\bf \mathcal{S}}$ is nondefective, using eigendecomposition, there is a non-singular matrix $\mathcal{W}$ and diagonal matrix $\bm{\Gamma}$ such that
\begin{equation}\label{eq:SU}
{\bf \mathcal{S}} = \mathcal{W}\bm{\Gamma}\mathcal{W}^{-1} \;,
\end{equation}
where
\begin{equation}
\bm{\Gamma} = \text{diag}\left(-2, -2, \frac{a+\sqrt{4+a^2}}{a}, \frac{a-\sqrt{4+a^2}}{a}\right)\;.
\end{equation}
The largest and smallest singular values of ${\bf \mathcal{S}}$ are $\sigma_1({\bf \mathcal{S}})=\frac{a+\sqrt{4+a^2}}{a}$ and $\sigma_4({\bf \mathcal{S}})=\min\Big(2,\frac{\sqrt{4+a^2}-a}{a}\Big)$ respectively. 
Based on~\eqref{eq:asa}, we have
\begin{equation}\label{eq:sig1-4}
 \sigma_1(\M)\leq \sigma_1({\bf \mathcal{S}})\, \sigma_1(\A\A^T) \;,
\end{equation}
\begin{equation}\label{eq:sig4-4}
 \sigma_4(\M)\geq \sigma_4({\bf \mathcal{S}})\, \sigma_4(\A\A^T) \;.
\end{equation}
Therefore, to bound  $\kappa_4(\M) = \sigma_1(\M)/\sigma_4(\M)$, we need to derive the bound for $\sigma_1(\A\A^T)$ and $\sigma_4(\A\A^T)$. 
Assuming a uniform distribution of the microphones on the circular table, we have the following distribution for $\rho$
\begin{equation}
P_{\rho}(\rho) = \frac{2\rho}{a^2} \quad \quad \quad\text{for}\quad   0\leq\rho\leq a \;.
\end{equation}
Therefore, the expectation of the matrix $\A^T\A$ is 
\begin{eqnarray}
	\E[\A^T\A] = \begin{bmatrix}Na^2/4&0&0&Na^3/8\\0&Na^4/4&0&0\\0&0&Na^4/4&0\\Na^3/8&0&0&7Na^4/48\end{bmatrix} \;.
\end{eqnarray}
Hence, the largest and smallest singular values of $\E[\A^T\A]$ are $N\sigma_{\max}(a)$ and $N\sigma_{\min}(a)$\,respectively with $\sigma_{\max}(a)$ and $\sigma_{\min}(a)$ independent of $N$. Moreover, $\sigma_i(\cdot)$ is a Lipschitz continuous function of its arguments and based on the Chernoff bound~\cite{trop12}, we get
\begin{eqnarray}
	\prob( \sigma_1(\A\A^T) > 2N\sigma_{\max}(a)) \leq e^{-\mathcal{C}^\prime N}\label{eq:2norm1}\;, \\ 
	\prob( \sigma_1(\A\A^T) < (1/2)N\sigma_{\max}(a)) \leq e^{-\mathcal{C}^\prime N}\label{eq:2norm2}\;, \\ 
	\prob( \sigma_4(\A\A^T) < (1/2)N\sigma_{\min}(a)) \leq e^{-\mathcal{C}^\prime N}\label{eq:2norm3}\;,
\end{eqnarray}
for a constant $\mathcal{C}^\prime$. Hence, with high probability, based on relations \eqref{eq:sig1-4}, \eqref{eq:sig4-4}, \eqref{eq:2norm1} and \eqref{eq:2norm3}, we have 
\begin{equation}
 \kappa_4(\M)\leq \frac{4\sigma_{\max}( a)\,\sigma_1({\bf \mathcal{S}})}{\sigma_{\min}(a)\,\sigma_4({\bf \mathcal{S}})} = f_{\kappa_4}(a)\;.
\end{equation}
This bound is independent of $N$. 

In the next step, we have to bound $\mu_1$ and $\mu_2$. The rank of matrix $\A$ is $\eta$, therefore there are matrices $\B\in \R^{\eta \times \eta}$ and $\V\in \R^{ N \times \eta}$ such that $\A = \V\B^T$ and $\V^T\V=N\eye$. Given $\M=\U \Sigma \U^T$ and~\eqref{eq:asa}, we have $\Sigma = \Q^T\B^T{\bf \mathcal{S}}\B\Q$ and $\U = \V\Q$ for an orthogonal matrix $\Q$. 
To show the incoherence property $\mathcal{A}_1$, we show that 
\begin{equation}
	\|\V_{i.}\|^2\leq \eta \, \mu_1 \quad \quad \forall \;  i\in[N]\;,
\end{equation} 
% \begin{eqnarray*}
% 	\|\V_i\|^2\leq \frac{4\mu}{n}\; .
% \end{eqnarray*} 
where $\V_{i.}$ denotes the transpose of $i^{\text{th}}$ row of the corresponding matrix. For $\eta = 4$, since $\V_{i.}=\B^{-1}\A_{i.}$, we have $\|\V_{i.}\|^2 \leq \sigma_4(\B)^{-2}\|\A_{i.}\|^2$ and $\sigma_4(\A)= \sqrt{N}\,\sigma_4(\B) $, therefore
\begin{equation}\label{eq:vi.}
 \|\V_{i.}\|^2 \leq \sigma_4(\A)^{-2}\|\A_{i.}\|^2\, N\;.
\end{equation}
Moreover, $\|\A_{i.}\|^2=a^2/4+\rho_i^2+(-a^2/4+\rho_i^2)^2 \leq 5a^2/4+ 9a^4/16$.
Defining
\begin{equation}
  f_{\mu_1}(a) =  \frac{5a^2/2+ 9a^4/8}{\sigma_{\min}(a)}\;,
\end{equation}
and based on \eqref{eq:2norm3} and \eqref{eq:vi.}, with high probability we have 
\begin{equation}\label{eq:incoprf1}
	\|\U_{i.}\|^2 \leq f_{\mu_1}(a) \quad \quad \quad \quad \forall \;  i\in[N]\;.
\end{equation}
Therefore, the incoherence property $\mathcal{A}_1$ for $\mu_1 = f_{\mu_1}(a)/\eta $ is correct; that is independent of $N$.

To prove the incoherence property $\mathcal{A}_2$, it is enough to prove that $\big| \M_{ij}/\sigma_1(\M) \big|\leq\sqrt{\eta} \; \mu_2/N$ for all $i,j \, \in [N]$.   
The maximum value of $\M_{ij}$ is $4a^2$ and based on \eqref{eq:sig4-4} and \eqref{eq:2norm3} we have
\begin{equation}
 \sigma_1(\M)\geq \sigma_4(\M)\geq \frac{1}{2} N\,\sigma_{\min}(a)\, \sigma_{4}({\bf \mathcal{S}})\;,
\end{equation}
Defining  $f_{\mu_2}(a)= 8a^2/\sigma_{\min}(a)\, \sigma_{4}({\bf \mathcal{S}})$, we have  
\begin{equation}
 \big| \M_{ij}/\sigma_1(\M) \big|\leq \frac{f_{\mu_2}(a)}{N} \quad \quad \quad \forall \,  i,j \; \in [N]\;.
\end{equation}
Therefore, the incoherence property $\mathcal{A}_2$ for $\mu_2 = f_{\mu_2}(a)/ \sqrt{\eta}$ is correct; that is independent of $N$. Since $\kappa_4(\M)$, $\mu_1$ and $\mu_2$ are bounded independent of $N$, matrix $\M$ is $(\mu_1,\mu_2)$-incoherent and the inequalities~\eqref{eq:if1} and~\eqref{eq:if11} are correct. 

Further,~\eqref{eq:if2} holds with high probability, if the right-hand side of~\eqref{eq:mainthm1} is less than $C_3 \,\sigma_{\min}(a) \, \sigma_4(\bf \mathcal{S})$, since based on \eqref{eq:2norm3}, $\frac{\sigma_{\eta}(\M)}{N} \geq \frac{1}{2} \sigma_{\min}(a) \sigma_4({\bf \mathcal{S}})$. This finishes the proof of Theorem \ref{thm:main1}. 

{\small \myqed}

{\color{blue} The theoretical analysis elaborated in this section, elucidates a link between the performance of microphone array calibration and the number of microphones, noise level and the ratio of missing pairwise distances. In Section~\ref{sec:experim}, thorough evaluations are conducted that demonstrate these theoretical insights.} Furthermore, The theoretical error bounds of ad~hoc microphone calibration established above corresponds to the classic matrix completion algorithm. We will extend the mathematical results to the completion of Euclidean distance matrices incorporating the Cadzow and EDM projections through the experiments. As we will see in Section~\ref{sec:experim}, this bound is not tight for the Cadzow projection and the Euclidean distance matrix completion algorithm as we achieve better results than matrix completion for microphone array calibration.

%%%%%%%%%%%%%%%%%%%%%%%%%%%%%%%%%%%%%%%%%%%%%%%%%%%%%%%%%%%%%%%%%%%%%%%%%%%%%%%%%%%%%%%%%%%%
\section{Related Methods}
\label{sec:topology}
%%%%%%%%%%%%%%%%%%%%%%%%%%%%%%%%%%%%%%%%%%%%%%%%%%%%%%%%%%%%%%%%%%%%%%%%%%%%%%%%%%%%%%%%%%%%
The objective is to extract the relative (up to a rigid transformation) microphone positions $\mathbi{x}_i,\, i \in \{1,\ldots ,N\}$ from the measurements of pairwise distances. Some of the state-of-the-art methods to achieve this goal are (1) Multi-Dimensional Scaling (MDS)~\cite{MDS_MAP}, (2) Semi-Definite Programming (SDP)~\cite{biswas06} and S-Stress~\cite{book-MDS} discussed briefly in the following sections. We refer the reader to the references for further details.

%%%%%%%%%%%%%%%%%%%%%%%%%%%%%%%%%%%%%%%%%%%%%%%%%%%%%%%%%%%%%%%%%%%%%%%%%%%%%%%%%%%%%%%%%%%%
\subsection{Classic Multi-Dimensional Scaling Algorithm}
\label{sec:MDS} 
%%%%%%%%%%%%%%%%%%%%%%%%%%%%%%%%%%%%%%%%%%%%%%%%%%%%%%%%%%%%%%%%%%%%%%%%%%%%%%%%%%%%%%%%%%%%
MDS refers to a set of statistical techniques used in finding the configuration of objects in a low dimensional space such that the measured pairwise distances are preserved \cite{MDS}. Given a distance matrix, finding the relative microphone positions is achieved by MDSLocalize~\cite{book-MDS}. In the ideal case where matrix $\M$ is complete and noiseless, this algorithm outputs the relative positions of the microphones. At the first step, a double centering transformation is applied to $\M$ to subtract the row and column means of the distance matrix via $\varXi( \M)= \frac{-1}{2}\, \J \, \M \, \J$ where $\J = \I_N - 1/N \ones_N \ones_N^T$. The $\zeta$ largest eigenvalues and the corresponding eigenvectors of $\varXi(\M)$ denoted by ${\bm \varPi}_+$ and $\U_+$ are calculated and the microphone positions are obtained as $\X = \U_+ \sqrt{{\bm \varPi}_+}$.

%  in the desired dimension through the steps summarized as follows
% \begin{itemize}
% \renewcommand{\labelitemi}{$\diamond$}\itemsep0em
% \item Double centering $\M$ via $\varXi( \M)= \frac{-1}{2}\, \J \, \M \, \J$.
% \item Eigenvalue decomposition of $\varXi(\M)$ as $\U {\bm \varPi} \U^T$. 
% \item Extracting the $\zeta$ largest eigenvalues and the corresponding eigenvectors denoted by ${\bm \varPi}_+$ and $\U_+$. 
% \item The microphone positions are obtained as $\X = \U_+ {\bm \varPi}_+ $.
% \end{itemize}

In a real scenario of missing distances, a modification called MDS-MAP~\cite{MDS_MAP} computes the shortest paths between all pairs of nodes in the region of consideration. The shortest path between microphones $i$ and $j$ is defined as the path between two nodes such that the sum of the estimated distance measures of its constituent edges is minimized. By approximating the missing distances with the shortest path and constructing the distance matrix, classical MDS is applied to estimate the microphone array geometry. 

%%%%%%%%%%%%%%%%%%%%%%%%%%%%%%%%%%%%%%%%%%%%%%%%%%%%%%%%%%%%%%%%%%%%%%%%%%%%%%%%%%%%%%%%%%%%
\subsection{Semidefinite Programming}
%%%%%%%%%%%%%%%%%%%%%%%%%%%%%%%%%%%%%%%%%%%%%%%%%%%%%%%%%%%%%%%%%%%%%%%%%%%%%%%%%%%%%%%%%%%%
Another efficient method that can be used for calibration is the semidefinite programming approach formulated as
\begin{equation}
\label{eq:sdp}
\hX = \argmin_{\X} \sum_{(i,j) \in E}  w_{ij} \left| \norm{\bm{x}_i - \bm{x}_j}^2 - \td_{ij}^2 \right|\;,
\end{equation}
where $ w_{ij} $ shows the reliability measure on the estimated pairwise distances. The basis vectors in Euclidean space $\R^N$ are denoted by $\{u_1, u_2, \cdots , u_N\}$. The optimization expressed in equation \eqref{eq:sdp} is not convex but can be relaxed as a convex minimization via
\begin{equation}
\begin{split}
 \min_{\X,\Y} \sum_{(i,j) \in E}  w_{ij} \left| (u_i-u_j)^T [\Y, \X ;\X^T,\I_{\zeta}](u_i-u_j)^T - \td_{ij}^2 \right| \\
 \text{subject   to} \quad [\Y, \X ;\X^T,\I_{\zeta}]\succeq 0,\quad \norm{\X^T \ones_N}=0\quad \quad
\end{split}
\end{equation}
where $\Y_{N \times N}$ is a positive semidefinite matrix and $\succeq$ is a generalized matrix inequality on the positive semidefinite cone \cite{book-boyd}.
To further increase the accuracy, a gradient decent is applied on the output of SDP minimization~\cite{biswas06}.

%%%%%%%%%%%%%%%%%%%%%%%%%%%%%%%%%%%%%%%%%%%%%%%%%%%%%%%%%%%%%%%%%%%%%%%%%%%%%%%%%%%%%%%%%%%%
\subsection{Algebraic S-Stress Method}
\label{sec:ss}
%%%%%%%%%%%%%%%%%%%%%%%%%%%%%%%%%%%%%%%%%%%%%%%%%%%%%%%%%%%%%%%%%%%%%%%%%%%%%%%%%%%%%%%%%%%%
The s-stress method for calibration extracts the topology of the ad~hoc network by optimizing the cost function stated as
\begin{equation}
\label{ss}
 \hX = \argmin_{\X} \sum_{(i,j) \in E}  w_{ij} \left( \norm{\bm{x}_i - \bm{x}_j}^2 - \td_{ij}^2 \right)^2\;.
\end{equation}
The reliability measure $w_{ij}$ controls the least square regression stated in equation \eqref{ss} which can be set according to the measure of $\td_{ij}$. If $w_{ij} = \td_{ij}^{-2}$, we have \textit{elastic scaling} that gives importance to large and small distances. If $w_{ij}=1$, large distances are given more importance than the small distances. In general, incorporation of $w_{ij} = \td_{ij}^\alpha,\, \alpha \in \{ ...,-2,-1,0,1,2,...\}$ yields different loss functions and depending on the structure of the problem, one of them may work better than the other~\cite{buja02}.

\section{Experimental Analysis}
\label{sec:experim}

%%%%%%%%%%%%%%%%%%%%%%%%%%%%%%%%%%%%%%%%%%%%%%%%%%%%%%%%%%%%%%%%%%%%%%%%%%%%%%%%%%%%%%%%%%%%%%%%%%%%
\subsection{A-priori Expectations}
%%%%%%%%%%%%%%%%%%%%%%%%%%%%%%%%%%%%%%%%%%%%%%%%%%%%%%%%%%%%%%%%%%%%%%%%%%%%%%%%%%%%%%%%%%%%%%%%%%%%
The simplest method that we discussed is the classical MDS algorithm. This method assumes that all the pairwise distances are known and in the case of missing entries and noise, it does not minimize a meaningful utility function. An extension of this method is MDS-MAP which replaces the missing distances with the shortest path. In many scenarios, this is considered as a coarse approximation of the true distances. 

The SDP-based method on the other hand is known to perform fairly well with missing distance information. Together with its final gradient descent phase, has been shown to find good estimates of the location. However, since each distance information translates into a constraint in the semi-definite program, this approach is not scalable and becomes intractable for large sensor networks. 

The alternative approach is to minimize the non-convex s-stress function. Although it is known to perform well in many conditions, in the case of missing distances, one cannot eliminate the possibility of falling into local minima using this approach. 

The approach that we proposed in this paper exploits a matrix completion algorithm to recover the missing distances considering the low-rank as well as Euclidean properties of the distance matrix. The classic matrix completion does not take into account the EDM properties. By integrating the Cadzow projection, the estimated matrix has partial EDM properties, and hence we expect better reconstruction results. Further, by incorporating the full EDM structure, we achieve a Euclidean distance matrix completion algorithm and expect more fidelity in the reconstruction performance. 
{\color{blue}In this section, we present thorough evaluation of ad~hoc microphone array calibration on simulated setups and real data recordings.}

{\color{blue}
%%%%%%%%%%%%%%%%%%%%%%%%%%%%%%%%%%%%%%%%%%%%%%%%%%%%%%%%%%%%%%%%%%%%%%%%%%%%%%%%%%%%%%%%%%
%%%%%%%%%%%%%%%%%%%%%%%%%%%%%%%%%%%%%%%%%%%%%%%%%%%%%%%%%%%%%%%%%%%%%%%%%%%%%%%%%%%%%%%%%%
%%%%%%%%%%%%%%%%%%%%%%%%%%%%%%%%%%%%%%%%%%%%%%%%%%%%%%%%%%%%%%%%%%%%%%%%%%%%%%%%%%%%%%%%%%
\subsection{Simulated Data Evaluations}
%%%%%%%%%%%%%%%%%%%%%%%%%%%%%%%%%%%%%%%%%%%%%%%%%%%%%%%%%%%%%%%%%%%%%%%%%%%%%%%%%%%%%%%%%%
The simulated experiments are conducted to evaluate the performance of the proposed method and compare and contrast it against the state-of-the-art alternative approaches in different scenarios with varying number of microphones, magnitude of the pairwise distance measurements errors, percentage of missing distances as well as jitter. 

{\color{purple}The presented evaluation relies on a local connectivity assumption in pairwise distance measurements. We do not assume a particular (e.g. diffuse noise) model for pairwise distance estimation and the conclusions of this section hold for a general ad~hoc array calibration framework where the pairwise distances may be provided by any other means meeting the local connectivity assumption.}

%%%%%%%%%%%%%%%%%%%%%%%%%%%%%%%%%%%%%%%%%%%%%%%%%%%%%%%%%%%%%%%%%%%%%%%%%%%%%%%%%%%%%%%%%%
\subsubsection{Performance for Different Numbers of Microphones}\label{sec:num_mic}
%%%%%%%%%%%%%%%%%%%%%%%%%%%%%%%%%%%%%%%%%%%%%%%%%%%%%%%%%%%%%%%%%%%%%%%%%%%%%%%%%%%%%%%%%%
In this section, we present the performance of ad~hoc array calibration when the number of microphones varies from 15 to 200. 
The microphones are uniformly distributed on a disc of diameter 19 m. The maximum pairwise distance that can be measured is 7.5 m. In addition, 5\% of the distances are assumed to be randomly missing. Hence, the total missing entries vary from 42\% to 60\%. The standard deviation of the noise on measured distances (expressed through~\eqref{eq:output}-\eqref{eq:def_Up}) between two microphones $i$ and $j$ is $\varsigma\, d_{ij}$ where $\varsigma = 0.0167$; the dependency of the noise level on the distance is due to the limitation of the diffuse noise coherence model for pairwise distance estimation as elaborated in~\cite{Taghizadeh2014242}. 

The results for each number of microphones are averaged over 500 random configurations. The calibration error is quantified using the metric defined in~\eqref{distance}. Furthermore, the absolute position of the microphones is estimated using the nonlinear optimization method~\cite{SDP} and the mean position error as defined in~\eqref{eq:pos-err} over all configurations is evaluated. Figures~\ref{fig:num_mic} and~\ref{fig:num_mic_crb} illustrate the results; the error bars are shown for one standard deviation from the mean estimates. The Cram\'er rao bound (CRB) is quantified using the method elaborated in~\cite{raykar2005,crocco2012bilinear}. %,crocco2012bilinear

The results show that the performance improves as the number of microphones increases. This observation is inline with the theoretical analysis provided in Section~\ref{sec:main}. The best results are achieved by the proposed E-MC$^2$ algorithm as it confines the search space to the Euclidean space through iterative EDM projections. We can see that for the number of microphones above 45, the error in position estimation is less than 6.2 cm and it reduces to 2.2 cm for 200 microphones. Although the mathematical proof of the unbiasedness of the proposed estimator is not achieved in this paper, we empirically found no evidence of bias. Therefore, CRB provides a reasonable benchmark for our evaluation. 
           
% %%%%%%%%%%%%%%%%%%%%%%   FIGURE 3  %%%%%%%%%%%%%%%%%%%%%%%%%%
% \begin{figure}[h]
%  \begin{center}
%   \includegraphics[width=1\linewidth]{num_mic-crop}
% \end{center}
% \caption{{\color{blue}Calibration error (logarithmic scale) as defined in~\eqref{distance} versus the number of microphones. The standard deviation of noise on measured distances is $\varsigma\, d_{ij}$ where $\varsigma = 0.0167$. The error bars correspond to one standard deviation from the mean estimates.}}
% \label{fig:num_mic}
% \end{figure}
% %%%%%%%%%%%%%%%%%%%%%%%%%%%%%%%%%%%%%%%%%%%%%%%%%%%%%%%%%%%%%
% 
% %%%%%%%%%%%  FIGURE 4 %%%%%%%%%%%%%%%%%%%%%%%%%%%%%%%%
% \begin{figure}[h]
%  \begin{center}
%   \includegraphics[width=1\linewidth]{num_mic_crb-crop}
% \end{center}
% \caption{{\color{blue}Mean position error (logarithmic scale) as defined in~\eqref{eq:pos-err} versus the number of microphones. The standard deviation of the noise on measured distances is $\varsigma\, d_{ij}$ where $\varsigma = 0.0167$. The error bars correspond to one standard deviation from the mean estimates.}}
% \label{fig:num_mic_crb}
% \end{figure}
% %%%%%%%%%%%%%%%%%%%%%%%%%%%%%%%%%%%%%%%%%%%%%%%%%%%%%

%%%%%%%%%%%%%%%%%%%%%%%%%%%%%%%%%%%%%%%%%%%%%%%%%%%%%%%%%%%%%%%%%%%%%%%%%%%%%%%%%%%%%%%%%%
\subsubsection{Performance for Different Noise Levels}
%%%%%%%%%%%%%%%%%%%%%%%%%%%%%%%%%%%%%%%%%%%%%%%%%%%%%%%%%%%%%%%%%%%%%%%%%%%%%%%%%%%%%%%%%%
To evaluate the effect of noise on calibration performance, similar (500) configurations of 45 microphones as generated in Section~\ref{sec:num_mic} are simulated. The level of white Gaussian noise added to the measured pairwise distance $d_{ij}$ are varying as $\varsigma\,d_{ij}$ where $\varsigma = \{0.0056,\hdots,0.1\}$. Figures~\ref{fig:snr_sim} and~\ref{fig:snr_crb_sim} illustrate the results. We can see that the performance improves as the noise level gets smaller. 

Based on the theoretical analysis of Section~\ref{sec:main} as expressed in~\eqref{eq:mainthm1}, a linear relationship between the calibration error of matrix completion and $\varsigma$ is expected. The empirical observations are in line with this theoretical insight. As depicted in Figure~\ref{fig:snr_sim}, for $\varsigma < 0.0167$, the second term in~\eqref{eq:mainthm1} is getting too small so the first term becomes dominant as the slope of the error reduction is reduced. 

% %%%%%%%%%%%%%%%%%%%%%%   FIGURE 5  %%%%%%%%%%%%%%%%%%%%%%%%%%
% \begin{figure}[h]
%  \begin{center}
%   \includegraphics[width=1\linewidth]{snr-crop}
% \end{center}
% \caption{{\color{blue}Calibration error (logarithmic scale) as quantified in~\eqref{distance} versus $\varsigma$. The error bars correspond to one standard deviation from the mean estimates.}}
% \label{fig:snr_sim}
% \end{figure}
% %%%%%%%%%%%%%%%%%%%%%%%%%%%%%%%%%%%%%%%%%%%%%%%%%%%%%%%%%%%%%

% % %%%%%%%%%%%%%%%%%%%%%%   FIGURE 6  %%%%%%%%%%%%%%%%%%%%%%%%%%
% \begin{figure}[h]
%  \begin{center}
%   \includegraphics[width=1\linewidth]{snr_crb-crop}
% \end{center}
% \caption{{\color{blue}Mean position error (logarithmic scale) as defined in~\eqref{eq:pos-err} versus $\varsigma$. The error bars correspond to one standard deviation from the mean estimates.}}
% \label{fig:snr_crb_sim}
% \end{figure}
% % %%%%%%%%%%%%%%%%%%%%%%%%%%%%%%%%%%%%%%%%%%%%%%%%%%%%%%%%%%%%%

%%%%%%%%%%%%%%%%%%%%%%%%%%%%%%%%%%%%%%%%%%%%%%%%%%%%%%%%%%%%%%%%%%%%%%%%%%%%%%%%%%%%%%%%%%
\subsubsection{Performance for Different Missing Ratios}
%%%%%%%%%%%%%%%%%%%%%%%%%%%%%%%%%%%%%%%%%%%%%%%%%%%%%%%%%%%%%%%%%%%%%%%%%%%%%%%%%%%%%%%%%%
To study the sensitivity of the proposed algorithm to different levels of missing distances, a cubic room of unit dimensions ($1 \times 1 \times 1$ m$^3$) is simulated and 60 microphones are distributed uniformly at random positions. 300 random configurations are generated and the average mean position error is evaluated. As an alternative approach, the self-calibration method proposed by Crocco et al.~\cite{crocco2012bilinear} is implemented considering 30 sources and 30 sensors {\color{purple}(thus 60 nodes in total). It may be noted that the number of nodes for calibration is equal for both approaches.} The distances between all source and microphone pairs are known.
Some of the distances are assumed to be missing at random. In addition, white Gaussian noise with standard deviation 0.02 m is added to the known distances. The simulated scenario mimics the evaluation setups of~\cite{crocco2012bilinear} and requires fixing the position of two microphones to derive the network position. 

Figure~\ref{fig:missing} illustrates the errors in position estimation for different ratios of missing distances. We can see that up to 50\% missing are effectively handled by the proposed algorithm. The rigorous analysis provided in Section~\ref{sec:main} requires that $Np \gg \text{log}N$ for the calibration error to be bounded; when the ratio of random missing entries is 60\% (i.e. $p=0.4$), we have $Np/\text{log}N = 5.85$ (violating the condition $\gg$) so the error in calibration is expected to increase significantly. The theoretical analysis is confirmed by this empirical observation. 

% %%%%%%%%%%%  FIGURE 7 %%%%%%%%%%%%%%%%%%%%%%%%%%%%%%%%
% \begin{figure}[h]
%  \begin{center}
%   \includegraphics[width=1\linewidth]{missing-crop}
% \end{center}
% \caption{{\color{blue}Mean position calibration error versus the ratio of missing pairwise distances for 30 sources and 30 microphones {\color{purple}(60 nodes in total)} considered in the self-calibration method~\cite{crocco2012bilinear} and 60 microphones used for the proposed E-MC$^2$ algorithm. The standard deviation of noise in pairwise distance estimation is 0.02.}}
% \label{fig:missing}
% \end{figure}
% %%%%%%%%%%%%%%%%%%%%%%%%%%%%%%%%%%%%%%%%%%%%%%%%%%%%%

%%%%%%%%%%%%%%%%%%%%%%%%%%%%%%%%%%%%%%%%%%%%%%%%%%%%%%%%%%%%%%%%%%%%%%%%%%%%%%%%%%%%%%%%%%
\subsubsection{Effect of Jitter on Calibration Performance}
%%%%%%%%%%%%%%%%%%%%%%%%%%%%%%%%%%%%%%%%%%%%%%%%%%%%%%%%%%%%%%%%%%%%%%%%%%%%%%%%%%%%%%%%%%
The study presented in this paper assumes that the microphones are synchronized prior to calibration. If a pilot signal at sampling frequency $f=$16 kHz is used for synchronization, the effect of jitter can be modeled by a uniform error in distance measures as $[\frac{-c}{2f},\frac{c}{2f}]$ where $c$ is the speed of sound and set to 340 m/s. Hence, we can model the jitter as an additional uniform noise on the distance measures within the range of $[-1.065,1.065]$ cm. 

The effect of jitter is evaluated for different levels of noise on the distances. The number of microphones is 45 distributed on a disc of diameter 19 m. 60\% of distances are missing consisting of 5\% random and 55\% structured. The experiments are repeated for 300 random configurations and the average calibration error and position estimation error are quantified. Figure~\ref{fig:jitter} illustrates the results. We can see that the effect of jitter on position estimation increases from 3 mm to 8 mm and its effect on calibration error increases from 0.01 m$^2$ to 0.09 m$^2$ as the distances are measured more accurately  (smaller $\varsigma$).    

% %%%%%%%%%%       FIGURE 8  %%%%%%%%%%%%%%%%%%%%%%%%%%%%%%%%
% \begin{figure}[h]
%  \begin{center}
%   \includegraphics[width=1\linewidth]{jitter-crop}
% \end{center}
% \caption{{\color{blue}Effect of jitter on E-MC$^2$ algorithm quantified in terms of (a) mean position error as defined in~\eqref{eq:pos-err} as well as (b) calibration error as defined in~\eqref{distance} versus $\varsigma$. The error bars correspond to one standard deviation from the mean estimates. The number of microphones is 45 and 60\% of the pairwise distances are missing.}}
% \label{fig:jitter}
% \end{figure}
% %%%%%%%%%%%%%%%%%%%%%%%%%%%%%%%%%%%%%%%%%%%%%%%%%%%%%%%%%%

%%%%%%%%%%%%%%%%%%%%%%%%%%%%%%%%%%%%%%%%%%%%%%%%%%%%%%%%%%%%%%%%%%%%%%%%%%%%%%%%%%%%%%%%%%
\subsubsection{Distributed Array Calibration}
%%%%%%%%%%%%%%%%%%%%%%%%%%%%%%%%%%%%%%%%%%%%%%%%%%%%%%%%%%%%%%%%%%%%%%%%%%%%%%%%%%%%%%%%%%
To further study the performance of the proposed approach for distributed array calibration, two scenarios are simulated. In the first scenario, a room of dimensions $11 \times 8 \times 5$~m$^3$ is considered which yields $d_{max} = 101$~cm~\cite{Taghizadeh2014242}~\footnote{{\color{purple}The maximum distance that can be estimated using the diffuse noise model depends on the size of the room and acoustic parameters. A linear relation between the maximum measurable distance and the room dimension has been shown rigorously~\cite{Taghizadeh2014242}. Nevertheless, application of the diffuse noise method for pairwise distance estimation is just an example use case of the proposed algorithm and many alternative approaches can be exploited.}}. {\color{purple}The reverberation time is about 430 ms.} Two sets of 9-channel circular uniform microphone array of diameter 20 cm are simulated where the center of both compact arrays are 1 m apart. 
In the second scenario, a room of dimensions $8 \times 5.5 \times 3.5$~m$^3$ is considered which yields $d_{max} = 73$ cm~\cite{taghiza13}. {\color{purple}The reverberation time is about 300 ms.} A circular 9-channel microphone array of diameter 20 cm located inside another 6-channel circular array of diameter 70 cm is simulated. 

The standard deviation of the noise on distance measures is $\varsigma\, d_{ij}$ where $\varsigma = 0.06$. There are no random missing entries and all of the missing distances are due to the limitation of the diffuse noise model in distance estimation thus around 25\% of the distances are missing in the first scenario (18-mic) and around 30\% of the distances are missing in the second scenario (15-mic). The results are listed in Table~\ref{tab:distributed}. {\color{purple}We can see that the positions are estimated with less than 1.6 cm error.} Furthermore, we repeated the same experiment 25 times and averaged the estimates of the positions. We can see that the error after averaging is noticeably reduced.   
\subsection{Real Data Evaluation}
%%%%%%%%%%%%%%%%%%%%%%%%%%%%%%%%%%%%%%%%%%%%%%%%%%%%%%%%%%%%%%%%%%%%%%%%%%%%%%%%%%%%%%%%%%%%
The real data recordings are collected at Idiap's smart meeting room~\cite{MONC}. 

%%%%%%%%%%%%%%%%%%%%%%%%%%%%%%%%%%%%%%%%%%%%%%%%%%%%%%%%%%%%%%%%%%%%%%%%%%%%%%%%%%%%%%%%%%
\subsubsection{Recording Set-up}
%%%%%%%%%%%%%%%%%%%%%%%%%%%%%%%%%%%%%%%%%%%%%%%%%%%%%%%%%%%%%%%%%%%%%%%%%%%%%%%%%%%%%%%%%%%%
We consider a scenario in which eleven microphones are located on a planar area: Eight of them are located on a circle with diameter $20$ cm and one microphone is at the center. There are two additional microphones at $70$ cm distance from the central microphone. The microphones are Sennheiser MKE-2-5-C omnidirectional miniature lapel type. {\color{purple} Although the recording setup for collecting data is regular, the uniform geometry of the microphone array provides no particular constraint. Hence, without loss of generality, we rely on this available setup to evaluate the performance of the proposed approach.}  

The floor of the room is covered with carpet and surrounded with plaster walls having two big windows. The enclosure is a $8 \times 5.5 \times 3.5$ m$^3$ rectangular room and it is moderately reverberant; {\color{purple}the reverberation time is about 300 ms}. It contains a centrally located $4.8 \times 1.2$ m$^2$ rectangular table. This scenario mimics the MONC database~\cite{MONC}. The sampling rate is $48$ kHz while the processing applied for microphone calibration is based on down-sampled signal of rate $16$ kHz to reduce the computational cost of pairwise distance estimation.   

%%%%%%%%%%%%%%%%%%%%%%%%%%%%%%%%%%%%%%%%%%%%%%%%%%%%%%%%%%%%%%%%%%%%%%%%%%%%%%%%%%%%%%%%%%%
\subsubsection{Pairwise Distance Estimation}
%%%%%%%%%%%%%%%%%%%%%%%%%%%%%%%%%%%%%%%%%%%%%%%%%%%%%%%%%%%%%%%%%%%%%%%%%%%%%%%%%%%%%%%%%%%%
In order to estimate the pairwise distances, we take two microphone signals of length $2.14$ s, frame them into short windows of length 1024 samples using a Tukey function (parameter = 0.25) and apply Fourier transform. For each frame, we compute the coherence function. The average of the coherence functions over 1000 frames are computed and used for estimation of the pairwise distance by fitting a $\sinc$ function as stated in~\eqref{eq:coherence} using the algorithm described in~\cite{taghiza13}. This algorithm is an improved version of the distance estimation using diffuse noise coherence model which enables a reasonable estimate up to $73$ cm. We empirically confirm that the distances beyond that are not reliably estimated so they are regarded as \emph{missing}. Thereby, the following entries of the Euclidean distance matrix are missing: $d_{10,11}, d_{1,10}, d_{7,10},$ $d_{8,10}, d_{5,11}, d_{6,11},$ $d_{7,11}$ (see Figure~\ref{fig:11mic_old}).

%%%%%%%%%%%%%%%%%%%%%%%%%%%%%%%%%%%%%%%%%%%%%%%%%%%%%%%%%%%%%%%%%%%%%%%%%%%%%%%%%%%%%%%%%%%
\subsubsection{Geometry Estimation }
%%%%%%%%%%%%%%%%%%%%%%%%%%%%%%%%%%%%%%%%%%%%%%%%%%%%%%%%%%%%%%%%%%%%%%%%%%%%%%%%%%%%%%%%%%%%
In the scenario described above, microphone calibration is achieved in two steps. First, all methods are used to find the nine close microphones in order to evaluate them for geometry estimation when we have all distances. The geometry of these microphones is fixed and used to calibrate the rest of the network. Figure~\ref{fig:9mic_old} demonstrates the results of MDS-MAP, SDP, s-stress and the proposed Euclidean distance matrix completion algorithm. The calibration error is quantified based on~\eqref{distance}. The best results are achieved by the proposed algorithm with error $5.85$ cm$^2$. The second place belongs to s-stress with error $6.14$ cm$^2$ followed by MDS-MAP and SDP with errors $8.13$ cm$^2$ and $8.63$ cm$^2$ respectively. 

% %%%%%%%%%%%%%%%%   FIGURE  11   %%%%%%%%%%%%%%%%%%%%%%%%
% \begin{figure}[h]
%  \begin{center}
%   \includegraphics[width=1\linewidth]{3fig9mic_jour2_old-crop}  % Done 3 !
%  \end{center}
% \caption{Calibration of the nine-element microphone array. The geometries are estimated using MDS-MAP, S-stress, SDP and the proposed Euclidean distance matrix completion algorithm, E-MC$^2$.}
% \label{fig:9mic_old}
% \end{figure}
% %%%%%%%%%%%%%%%%%%%%%%%%%%%%%%%%%%%%%%%%%%%%%%%%%%%%%%%%

Figure~\ref{fig:9mic_adv} provides a comparative illustration of the results of matrix completion (MC), MC+Cadzow (MC$^2$) and the proposed Euclidean distance matrix completion (E-MC$^2$) algorithm. 
We can see that MC$^2$ yields better result with error $7.68$ compared to MDS-MAP, SDP and MC, but worse than s-stress. The proposed E-MC$^2$ algorithm achieves the best performance.% Table~\ref{tab:all-methods9} summarizes all the results.   
 
% % %%%%%%%%%%%%%%%%   FIGURE  12   %%%%%%%%%%%%%%%%%%%%%%%%
% \begin{figure}[h]
%  \centering
%   \includegraphics[width=1\linewidth]{1fig9mic_jour2_adv-crop} % Done 1!
% \caption{{Calibration of the nine-element microphone array. The geometries are estimated using MC, MC+Cadzow (MC$^2$) and the proposed algorithm E-MC$^2$.}}
% \label{fig:9mic_adv}
% \end{figure}
% % %%%%%%%%%%%%%%%%%%%%%%%%%%%%%%%%%%%%%%%%%%%%%%%%%%%%%%%%

% \begin {table}[H]
% \centering
% \begin{tabular}{l*{3}{c}r}
% Methods                     & Calibration Error (cm$^2$)  &  Position Error (cm) \\
% \hline
% MDS-MAP                     &  8.13   &  0.78 \\
% SDP                         &  8.63   &  0.81 \\ 
% S-stress                    &  6.1    &  0.61 \\
% MC                          &  9.75   &  0.97 \\
% MC$^2$                      &  7.68   &  0.74 \\
% E-MC$^2$                    &  5.85   &  0.48 \\
% \end{tabular}
% \caption{Performance comparison of different methods for calibration of the nine-element microphone array. The calibration error is quantified based on~\eqref{distance} and {\color{blue}position error is defined in~\eqref{eq:pos-err}.}}  
% \label{tab:all-methods9}
% \end{table}

The scenario using eleven channels of microphones addresses the problem of having partial estimates of the distances for calibration of a microphone array. The experiments show that the proposed method offers the best estimation of the geometry as illustrated in Figure~\ref{fig:11mic_old} and \ref{fig:11mic_adv} with an error of $49.6$ cm$^2$. As we can see, the proposed Euclidean distance matrix completion algorithm achieves less than half the error of the best state-of-the-art alternative.

% %%%%%%%%%%%%%   FIGURE 10 %%%%%%%%%%%%%%%%%%%%%%%%%%%%
% \begin{figure}[h]
%  \begin{center}
%   \includegraphics[width=1\linewidth]{2fig11mic_jour2_old-crop}  % Done 2 !
% \end{center}
% \caption{Calibration of the eleven-element microphone array while several pairwise distances are missing.~The geometries are estimated using MDS-MAP, SDP, S-stress and the proposed proposed algorithm E-MC$^2$.}
% \label{fig:11mic_old}
% \end{figure}
% %%%%%%%%%%%%%%%%%%%%%%%%%%%%%%%%%%%%%%%%%%%%%%%%%%%%%%%

% % %%%%%%%%%%%%%   FIGURE 11 %%%%%%%%%%%%%%%%%%%%%%%%%%%%
% \begin{figure}[h]
% \centering
% \includegraphics[width=1\linewidth]{4fig11mic_jour2_adv-crop.pdf}  % Done 4 !
% \caption{Calibration of the eleven-element microphone array while several pairwise distances are missing.~The geometries are estimated using MC, MC+Cadzow (MC$^2$), and the proposed algorithm E-MC$^2$.}
% \label{fig:11mic_adv}
% \end{figure}
% %%%%%%%%%%%%%%%%%%%%%%%%%%%%%%%%%%%%%%%%%%%%%%%%%%%%%%%%%

The worst result belongs to MDS-MAP with error $434.4$ cm$^2$ because the shortest path is a poor estimation of missing entries. The s-stress and SDP search the Euclidean space corresponding to the feasible positions hence, their performance are more reasonable with errors $141$ cm$^2$ and $125$ cm$^2$. The advantage of being constrained to a physically possible search space or close to it is considered in extensions of matrix completion in MC+Cadzow (MC$^2$) and the proposed method (E-MC$^2$) and achieves the best performance. These experimental evaluation confirm the effectiveness of the proposed algorithm and demonstrate the hypothesis that incorporating the EDM properties in matrix completion algorithm enables calibration of microphone arrays from partial measurements of the pairwise distances.

{\color{blue} The theoretical analysis provided in Section~\ref{sec:main} elucidates a link between the calibration error and the number of microphones. To demonstrate this relation, a calibration of a 8-channel circular array when the distances are all measured is performed. In addition, an extra microphone {\color{purple}($\#$12)} is also included which is located with a symmetry to microphone $10$. Hence, $d_{12,11}, d_{10,12}, d_{3,12}, d_{4,12}, d_{5,12}$ are also missing. The calibration errors are listed in Table~\ref{tab:all-theory-cal}.

% % %%%%%%%%%%%%%%%%   FIGURE  13   %%%%%%%%%%%%%%%%%%%%%%%%
% \begin{table}%\footnotesize
% \centering
% \caption{{\color{blue} Calibration errors (cm$^2$) as defined in~\eqref{distance} for different methods of microphone array calibration.}}
% \begin{tabular}{@{}c*2{c}
%   c>{}c@{}}
% \toprule[1pt]
% & \multicolumn{2}{c}{\head{Known}}
% & \multicolumn{2}{c}{\head{Missing}}\\
% & \normal{\head{8-mic}} & \normal{\head{9-mic}}
% & \normal{\head{11-mic}} & \head{12-mic}\\
%   \cmidrule(lr){2-3}\cmidrule(l){4-5}
%   \multirow{1}{*}{MDS-MAP} & 9 & 8.13 & 434.4 & 472 \\
%   \cmidrule(lr){2-3}\cmidrule(lr){4-5}
%   \multirow{1}{*}{SDP} & 9.09 & 8.63 & 141 & 135 \\
%   \cmidrule(lr){2-3}\cmidrule(lr){4-5}
%   \multirow{1}{*}{S-Stress} & 6.86 & 6.14 & 125 & 95 \\
%   \cmidrule(lr){2-3}\cmidrule(lr){4-5}
%   \multirow{1}{*}{MC} & 10.6 & 9.75 & 133 & 115 \\
%   \cmidrule(lr){2-3}\cmidrule(lr){4-5}
%   \multirow{1}{*}{MC$^2$} & 9.2 & 7.68 & 119 & 52 \\
%   \cmidrule(lr){2-3}\cmidrule(lr){4-5}
%   \multirow{1}{*}{E-MC$^2$} & 6.5 & 5.85 & 49.6 & 46 \\
% \bottomrule[1pt]
% \end{tabular}\label{tab:all-theory-cal}
% \end{table}
% %%%%%%%%%%%%%%%%%%%%%%%%%%%%%%%%%%%%%%%%%%%%%%%%%%%%%%%%%%%%

Furthermore, in addition to the calibration error expressed in~\eqref{distance}, we apply the nonlinear optimization proposed in~\cite{SDP} to find the best match between $\hat{\X}$ and $\X$ by considering various rigid transformations and quantify the position error as
\begin{equation}\label{eq:pos-err}
 \frac{1}{N} \sum_{n=1}^{N}  \| \hat{{\bm x}}_{n} - {\bm x}_{n} \|_2\,.
\end{equation}
The position errors are listed in Table~\ref{tab:all-theory-pos}. The results show that considering further microphone improves the calibration performance which is in line with the theoretical analysis of Section~\ref{sec:main}.}

\section{Conclusions}\label{sec:conclusion}
We proposed a Euclidean distance matrix completion algorithm for calibration of ad~hoc microphone arrays from partially known pairwise distances. This approach exploits the low-rank property of the distance matrix and recovers the missing entries based on a matrix completion optimization scheme. To incorporate the properties of a Euclidean distance matrix, the estimated matrix at each iteration of the matrix completion is projected onto the EDM cone. Furthermore, we derived the theoretical bounds on the calibration error using matrix completion algorithm. The experimental evaluations conducted on real data recordings demonstrate that the proposed method outperforms the state-of-the-art techniques for ad~hoc array calibration. This study confirmed that exploiting the combination of the rank condition of EDMs, similarity in the measured distances, and iterative projection on the EDM cone leads to the best position reconstruction results. The proposed algorithm and the theoretical guarantees are applicable to the general framework of ad hoc sensor networks calibration.

% %%%%%%%%%%%%%%%%%%%%%%%%%%%%%%%%%%%%%%%%%%%%%%%
% \appendix
\vspace{+2mm}
% \textbf{\textit{Appendix 1.}}
\subsection*{\textbf{Appendix 1. Proof of Theorem~\ref{thm:main2}}}
\label{proof:thm2-app}
The goal is to find the bound of the norm of the squared distance matrix with missing entries according to structures indicated by $E$ and $S$. Based on~\eqref{eq:def_S} and~\eqref{eq:def_Pc}, we define matrix $\cE$ as 
\begin{equation}
 	\cE_{ij} = 
	\begin{cases}
		\; 1 & \text{if } (i,j) \in E\cap S\\
		\; 0 & \text{otherwise}
	\end{cases}
\end{equation}
Both $E$ and $S$ are symmetric matrices, hence~$\cE$ is also symmetric. Due to the physical setup, we know that $\Pc_E(\M)_{ij} \leq 4a^2$ for all $i, j \in  [N]$ and from the norm definition we have 
\begin{eqnarray}
	\|\Pc_E(\M^s)\|_2 &\leq& 4a^2 \max_{\|\textbf{h}\|=\| \bm \hbar \|=1} \sum_{i,j} |h_i|\, |\hbar_j|\, \cE_{ij} \label{eq:lem_Dbs2} =  4a^2 \|\cE\|_2\;, \nonumber
\end{eqnarray}
where $\textbf{h}=[h_1,h_2, . . . ,h_N]^T$ and ${\bm \hbar} = [\hbar_1, \hbar_2, . . .,\hbar_N]^T$ are right and left eigenvectors of matrix $\cE$. 
In order to bound $\| \cE \|_2$, we first define a binomial random variable vector $\bnu =[\nu_1, \nu_2,...,\nu_N]^T $ where
\begin{equation}\label{eq:Gresh}
 \nu_i = \sum_{j\in[N]} |\cE_{ij}|\;.
 \end{equation}
Based on the Gershgorin circle theorem we have $\|\cE\|_2  \leq \|\bnu \|_\infty$. Each entry in matrix $\cE\ $ is one with probability $p\,q$ where $q$ is the probability that the entry is included in structured missing entries or
\begin{equation}
q = \prob\{|\bm{x}_i - \bm{x}_j| \geq  d_\text{max}\}\;.
\end{equation}
Hence, we have
\begin{equation}
\label{eq:npq}
 \E[\nu_i] = Npq \;,
\end{equation}
For bounding $\E[\nu_i]$, it is necessary to bound $q$. Figure~\ref{fig:bounding}.I depicts the lowest probability of missing distances if the microphone location with respect to the edge of the circular table has a distance more than $d_\text{max}$ and Figure~\ref{fig:bounding}.II depicts the highest probability if the microphone is located right at the edge of the table. 

% %%%%%%%%%  FIGURE LAST    %%%%%%%%%%%%%%%%%%%%
% \begin{figure}[t]
%  \begin{center}
%   \includegraphics[width=1\linewidth]{Amin-Amax}
%  \end{center}
% \caption{Scenario corresponding to the (I) lower bound and (II) upper bound of the probability $q$ of structured missing distances.}
% \label{fig:bounding}
% \end{figure}
% %%%%%%%%%%%%%%%%%%%%%%%%%%%%%%%%%%%%%%%%%%%%%%%

The maximum of $d_\text{max}$ is $a$. We denote the upper bound and lower bound with $q_\text{max}(a,d_\text{max})$ and $q_\text{min}(a,d_\text{max})$ respectively, therefore 
\begin{equation}\label{eq:min-max}
	q_\text{min}(a,d_\text{max}) \leq q \leq  q_\text{max}(a,d_\text{max}) \;.
\end{equation}
As illustrated in  Figure~\ref{fig:bounding}. $q_\text{min}(a,d_\text{max}) = \max\{1- \big (\frac{d_\text{max}}{a}\big)^2,0 \}$ and $q_\text{max}(a,d_\text{max})= 1-\frac{B}{\pi a^2}$ where $B$ is the intersection area between the two circles. By computing $B$, we obtain 
\begin{equation}
\label{eq:Amax}
 q_{\text{max}}= 1-\frac{2 \gamma}{\pi}+\frac{1}{2\pi} \sin4\gamma+ \frac{2\xi^2}{\pi}[2\gamma+\sin2\gamma]-2\xi^2 \;,
\end{equation}
where $\xi= d_\text{max}/2a$ and $\gamma = \sin^{-1}\xi$.
Based on \eqref{eq:npq} and \eqref{eq:min-max} we have
\begin{equation}\label{eq:epqmin-max}
 Npq_\text{min}(a,d_\text{max}) \leq \mathbb E[\nu_i] \leq  Npq_\text{max}(a,d_\text{max}) \;.
\end{equation}
By applying the Chernoff bound to $\nu_i$ we have
\begin{equation}
\label{eq:chern}
 \prob \big(\nu_i > (1+\epsilon)\mathbb E[\nu_i] \big)\leq 2^{-(1+\epsilon) \mathbb E[\nu_i]}\;,
\end{equation}
where $\epsilon$ is an arbitrary positive constant. Therefore, based on \eqref{eq:epqmin-max} we have
\begin{equation}
\prob \big(\nu_i > (1+\epsilon)Np \, q_\text{max} \big)\leq 2^{-(1+\epsilon) Np \, q_\text{min}}\;.
\end{equation}
By applying the union bound we have
\begin{equation}
\prob \big(\max_{i\in [N]}\nu_i > (1+\epsilon)Np \, q_\text{max} \big)\leq 2^{-(1+\epsilon) Np \, q_\text{min}+\log_2 N}\;.
\end{equation}
We assume that $q_\text{min}$ and $q_\text{max}$ grow as $\mathcal{O}(\frac{\log_2N}{N})$; this assumption indicates that the ratio of the structured missing entries with respect to $N$ decreases as $N$ grows\footnote{This assumption can be dropped to achieve a tighter bound, but it increases the complexity of the proof.} or in other words, $d_{\text{max}}$ increases as the size of the network $N$ grows. Therefore, we have 
% $(\frac{d_\text{max}}{a}\big)^2$ grows as $1-\frac{\log_2 N}{N}$
%\footnote{This assumption means that as the number of microphones grow, we need to have better acoustic measurements, i.e., larger $d_\text{max}$.}. 
\begin{equation}
 \prob \big(\max_{i\in [N]}\nu_i > (1+\epsilon)Np \, q_\text{max} \big)\leq N^{-\theta}\;, \label{eq:boundend}
\end{equation}
where the positive parameter $\theta = (1+\epsilon)p-1$; by choosing $\epsilon \geq 4/p -1$, with probability greater than $1-N^{-3}$, we have
\begin{equation}
 \|\Pc_E(\M^s)\|_2 \leq 4a^2 \max_{i\in [N]} \nu_i\;,
\end{equation}
and based on \eqref{eq:boundend}
\begin{equation}
 \|\Pc_E(\M^s)\|_2 \leq 4 a^2  (1+\theta) q_\text{max}N\;.
\end{equation}
% By assumption rate of changing $q_\text{max}$ the same as the $q_\text{min}$~we have
% \begin{equation}
%  \|\Pc_E(\M^s)\|_2 \leq C_1 a^2  \log_2N
% \end{equation}
Therefore, we achieve   
\begin{equation}
\|\Pc_E(\M^s)\|_2 \leq C^{\prime \prime}_1 a^2  \log_2N\;.
\end{equation}

%T he value for $q_\text{max}$ for large $N$ based on \eqref{eq:Amax} is $1/3+ \sqrt{3}/(2\pi)$.   
\myqed
%%%%%%%%%%%%%%%%%%%%%%%%%%%%%%%%%%%%%%%%%%%%%%%%%%%%%%
\subsection*{\textbf{Appendix 2. Proof of Theorem~\ref{thm:main3}}}
\label{proof:thm3-app}
%%%%%%%%%%%%%%%%%%%%%%%%%%%%%%%%%%%%%%%%%%%%%%%%%%%
Based on the noise model described in Section~\ref{sec:noise},  $\Zsb_{ij}$ is obtained as
\begin{equation}
\label{eq:sub}
   \Zsb_{ij} = \, d_{ij}^2 {\bf\Upsilon}_{ij} \left(2+ {\bf\Upsilon}_{ij} \right) \;\approx \, 2d_{ij}^2 {\bf\Upsilon}_{ij} ,
\end{equation}
where $d_{ij} \leq d_\text{max}$ and based on concentration inequality for 1-Lipschitz function $\norm{.}$~on i.i.d random variables $\Pc_E(\Zsb)$ with zero mean and sub-Gaussian tail with parameter $4  \varsigma^2 d_{\max}^4$\eqref{eq:def_Up}, \eqref{eq:sub}~\cite{talagrand1996new} 
\begin{equation}
   \prob \left( \left| \; \norm{\Pc_E(\Zsb)}- \; \E \left(  \norm{\Pc_E(\Zsb)}    \right)  \; \right|  \; > t\; \right) \leq \exp \left(\frac{-t^2}{8 \; \varsigma^2 d_{\max}^4} \right).
\end{equation}

By setting $ t = 2d_{\max}^2\sqrt{6\varsigma^2 \log{N}}$  we have
\begin{equation}\label{eq:noise48}
 \norm{\Pc_E(\Zsb)}\leq \; \E \left(  \norm{\Pc_E(\Zsb)}\right) + 2d_{\max}^2\sqrt{6\varsigma^2 \log{N}}
\end{equation}
with probability bigger than $1-N^{-3}$.
So we need to extract bound for expectation of $\Pc_E(\Zsb)$ that has symmetric random enties. By using Theorem 1.1 from \cite{Seginer00},
\begin{equation}\label{eq:noise49}
 \E \left(  \norm{\Pc_E(\Zsb)}\right) \leq C_4 \;\E\left(\max_{j \in [N]} \norm{\Pc_E(\Zsb_{.j})} \right)
\end{equation}
Furthermore by using union bound and with apply Chernoff bound on the sum of independent random variables \cite{Keshavan10-2}
\begin{equation}\label{eq:noise50}
\E\left(\max_{j \in [N]} \norm{\Pc_E(\Zsb_{.j})}^2 \right) \leq C_5 d_{max}^4 \; \varsigma^2 pN
\end{equation}
Since
\begin{equation}\label{eq:noise51}
 \E\left(\max_{j \in [N]} \norm{\Pc_E(\Zsb_{.j})} \right) \leq  \sqrt {  \E\left(\max_{j \in [N]} \norm{\Pc_E(\Zsb_{.j})}^2 \right) }
\end{equation}
Base on relations \eqref{eq:noise49}, \eqref{eq:noise50} and \eqref{eq:noise51}
\begin{equation}\label{eq:noise52}
 \E \left(  \norm{\Pc_E(\Zsb)}\right) \leq C_6 d_{\max}^2 \varsigma \sqrt{pN}
\end{equation}
By using \eqref{eq:noise52} and \eqref{eq:noise48} for $pN\gg \log{N} $ we have
\begin{equation}\label{eq:noise53}
 \norm{\Pc_E(\Zsb)}\leq \; C^{\prime \prime}_2 d_{\max}^2 \varsigma \sqrt{pN}
\end{equation}
\myqed

% Based on the noise model described in Section~\ref{sec:noise}, the maximum entry of the matrix $\Zsb_{ij}$ is obtained as
% \begin{equation}
%   \max \Zsb_{ij} = \max_{i,j} \, d_{ij}^2 {\bf\Upsilon}_{ij} \left(2+ {\bf\Upsilon}_{ij} \right) \;,
% \end{equation}
% where $d_{ij} \leq d_\text{max}$ and based on~\eqref{eq:def_Up}, the value of $|{\bf\Upsilon}_{ij}|\left(2+|{\bf\Upsilon}_{ij}|\right)$ with probability greater than $1 - N^{-3}$ is less than $16\,\varsigma + 64\,\varsigma^2$ for a typical network of less than $N < 10^4$ microphones. Assuming a uniform distribution of the microphones, $\Zsb$ has $N\,(d_\text{max}/a)^2$ non zero entries. Based on the Gershgorin circle theorem~\cite{Scott-Gersh} we have
% \begin{equation}
%  \norm{\Zsb} \leq \max_i \sum_j |\Zsb_{ij}|\;.
% \end{equation}
% By applying $\Pc_E$ projection, with $C^{\prime \prime}_2(\varsigma) = 16\,\varsigma + 64\,\varsigma^2$~\footnote{The value of $\varsigma$ can be approximated with a constant for many types of rooms and estimated as less than $0.05$ through simulations.} we have
% \begin{equation}
% \label{eq:noiseb3}
%  \norm{\Pc_E(\Zsb)} \leq C^{\prime \prime}_2(\varsigma) \frac{d_\text{max}^4}{a^2}   p \, N \;.
% \end{equation} 
% 
% \myqed

% %%%%%%%%%%%%%%%%%%%%%%%%%%%%%%%%%%%%%%%%%%%%%%%

\section*{Acknowledgment}
This work was supported by the Swiss National Science Foundation under the National Center of Competence in Research (NCCR) on ``Interactive Multi-modal Information Management'' (IM2). The authors would like to acknowledge the anonymous reviewers for their comments and remarks to improve the quality and clarity of the manuscript.

% % \newpage
\bibliographystyle{elsarticle-num-names}%elsarticle-num-names  model3-num-names
\balance
\bibliography{strings11}
\newpage
%%%%%%%%%%%%%%%%%%%%%%%%%%%%%%%%%
\begin{table}[ht]
\caption{Summary of the notation.\vspace{0.2cm}}
\ra{0.8}
% \centering
\small
\begin{tabular}{@{}ll | ll@{}}
\toprule
Symbol & Meaning & Symbol & Meaning\\
\midrule
$N$ & number of microphones & $\D$ & complete noiseless distance matrix\\
$a$ & radius of the circular table on which microphones are distributed & $\M$ & squared distance matrix\\
$\varsigma$ & normalized standard deviation of noise & $\tM$ & noisy squared distance matrix\\	
$\Pc_E$ & projection into matrices with entries on index set $E$ & $\hM$ & estimated squared distance matrix\\
$\cP_e$ & projection to EDM cone & $\Z$ & noise matrix \\
$p$ & probability of having random missing entries  & $\M^E$ & observed matrix\\
$d_{max}$ & radius of the circle defining structured observed entries & $\X$ &positions matrix\\
$\Ms$ & distance matrix with observed entries on index set $S$ & $\hX$ &estimated positions matrix\\
\bottomrule
\end{tabular}
\label{tab:notations}
\end{table}
%%%%%%%%%%%%%%%%%%%%%%%%%%%%%%%%%
% 
% 
\newpage
%%%%%%%%%%  FIGURE 1   %%%%%%%%%%
\begin{figure}[ht]
\begin{center}
\includegraphics[width=\linewidth,height=.4\linewidth]{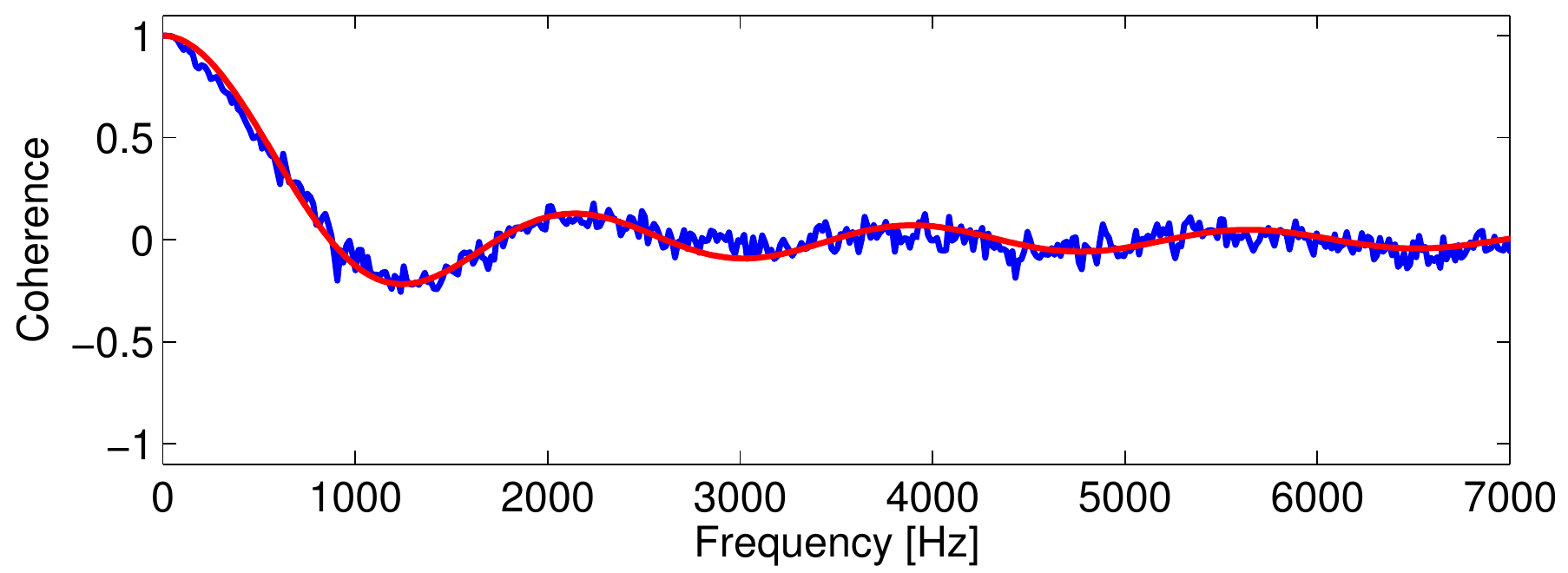} %synt_mic8_4_196-crop
\end{center}
\caption{{Coherence of the signal of two microphones at $d_{ij}=20\;$cm and the fitted $\sinc$ function using real data recordings.}}
\label{fig:sinc2}
\end{figure}
%%%%%%%%%%%%%%%%%%%%%%%%%%%%%%%%%
% 
% 
\newpage
%%%%%%%%%%%   FIGURE  2   %%%%%%%%%%%%%%%%%%%%%%%%%
\begin{figure}[ht]
\centering
\includegraphics[width=0.6\linewidth]{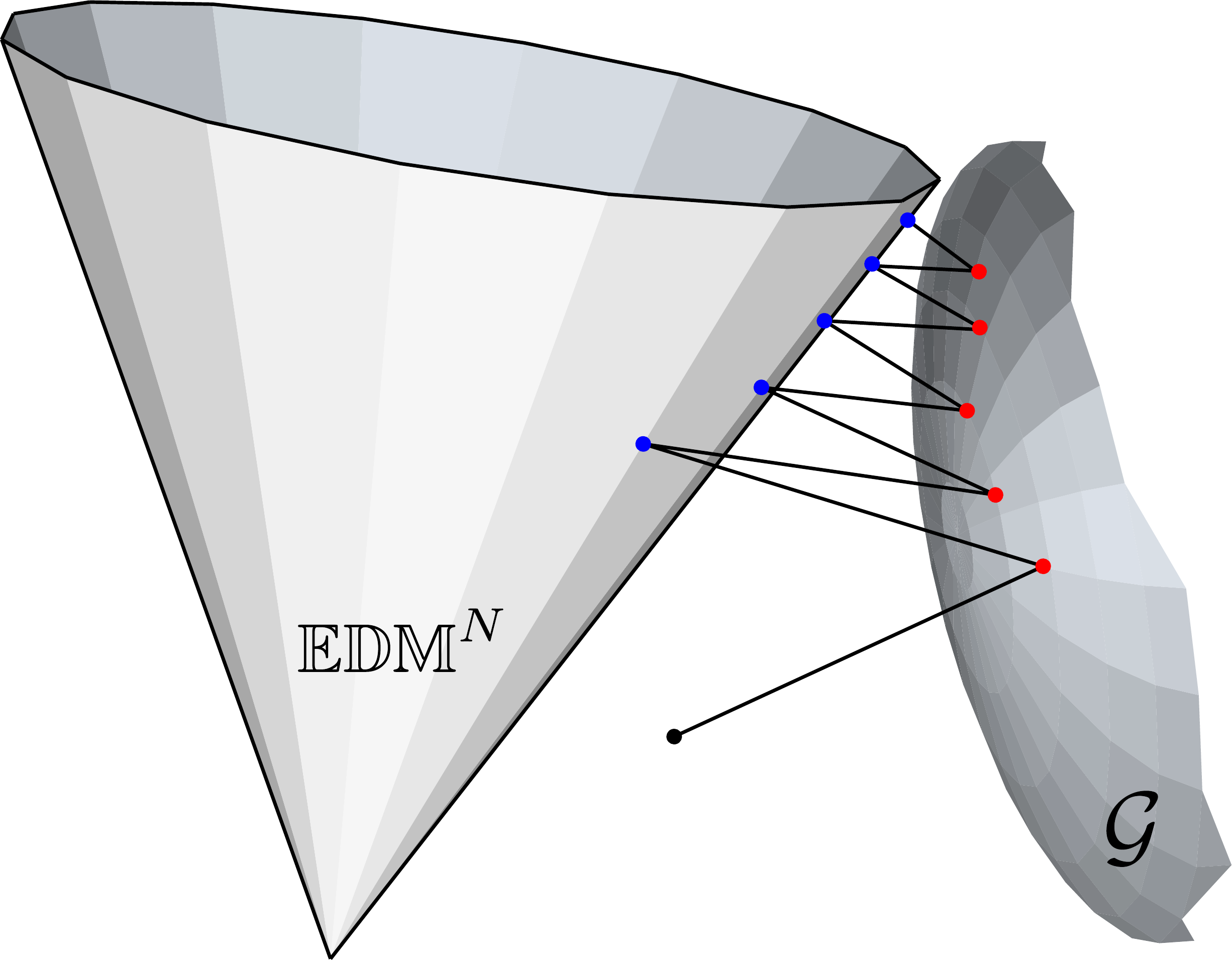}  %cone_etc
\caption{Matrix completion with projection onto the EDM cone. }
\label{fig:grassman}
\end{figure}
%%%%%%%%%%%%%%%%%%%%%%%%%%%%%%%%%%%%%%%%%%%%%%%%%%%
% 
\newpage
%%%%%%%%%%%%%%%%%%%%%%   FIGURE 3  %%%%%%%%%%%%%%%%%%%%%%%%%%
\begin{figure}[ht]
 \begin{center}
  \includegraphics[width=1\linewidth]{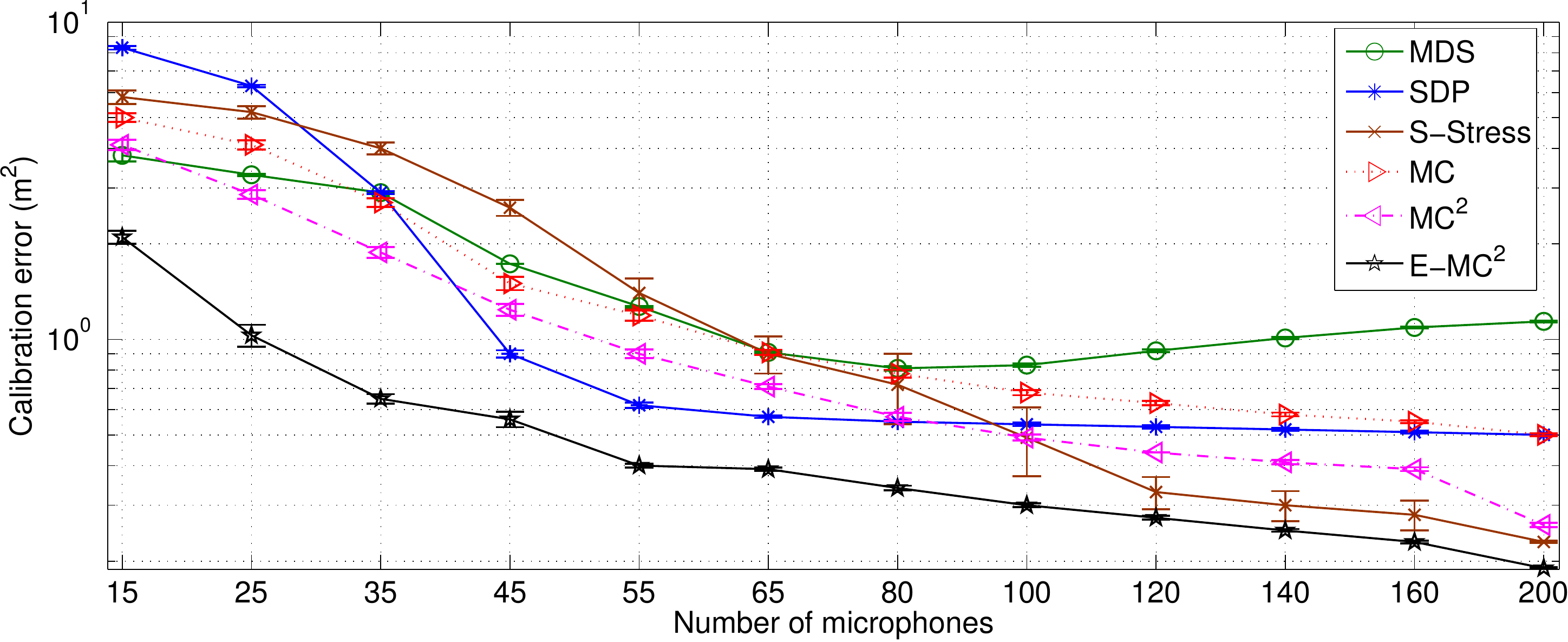}  %num_mic-crop
\end{center}
\caption{{\color{blue}Calibration error (logarithmic scale) as defined in~\eqref{distance} versus the number of microphones. The standard deviation of noise on measured distances is $\varsigma\, d_{ij}$ where $\varsigma = 0.0167$. The error bars correspond to one standard deviation from the mean estimates.}}
\label{fig:num_mic}
\end{figure}
%%%%%%%%%%%%%%%%%%%%%%%%%%%%%%%%%%%%%%%%%%%%%%%%%%%%%%%%%%%%%

\newpage
%%%%%%%%%%%  FIGURE 4 %%%%%%%%%%%%%%%%%%%%%%%%%%%%%%%%
\begin{figure}[ht]
 \begin{center}
  \includegraphics[width=1\linewidth]{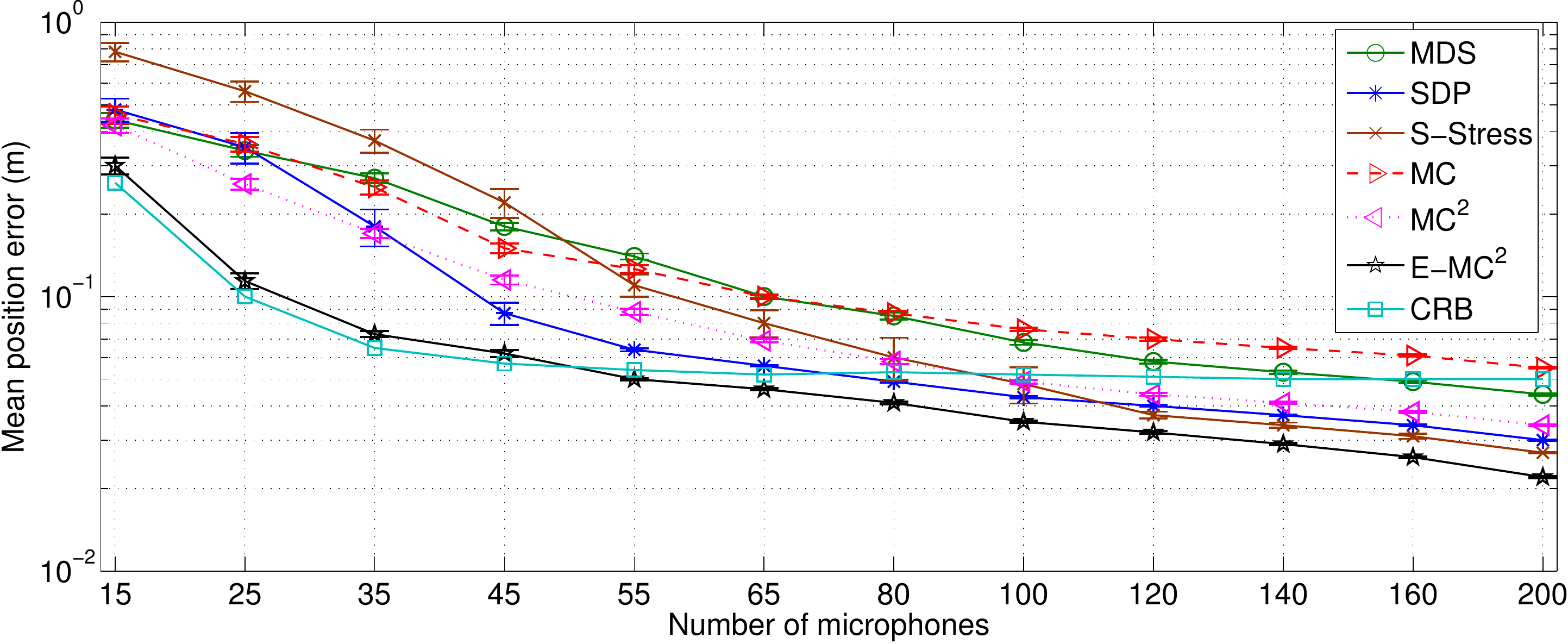}  %num_mic_crb-crop
\end{center}
\caption{{\color{blue}Mean position error (logarithmic scale) as defined in~\eqref{eq:pos-err} versus the number of microphones. The standard deviation of the noise on measured distances is $\varsigma\, d_{ij}$ where $\varsigma = 0.0167$. The error bars correspond to one standard deviation from the mean estimates.}}
\label{fig:num_mic_crb}
\end{figure}
%%%%%%%%%%%%%%%%%%%%%%%%%%%%%%%%%%%%%%%%%%%%%%%%%%%%%
% 
\newpage
%%%%%%%%%%%%%%%%%%%%%%   FIGURE 5  %%%%%%%%%%%%%%%%%%%%%%%%%%
\begin{figure}[ht]
 \begin{center}
  \includegraphics[width=1\linewidth]{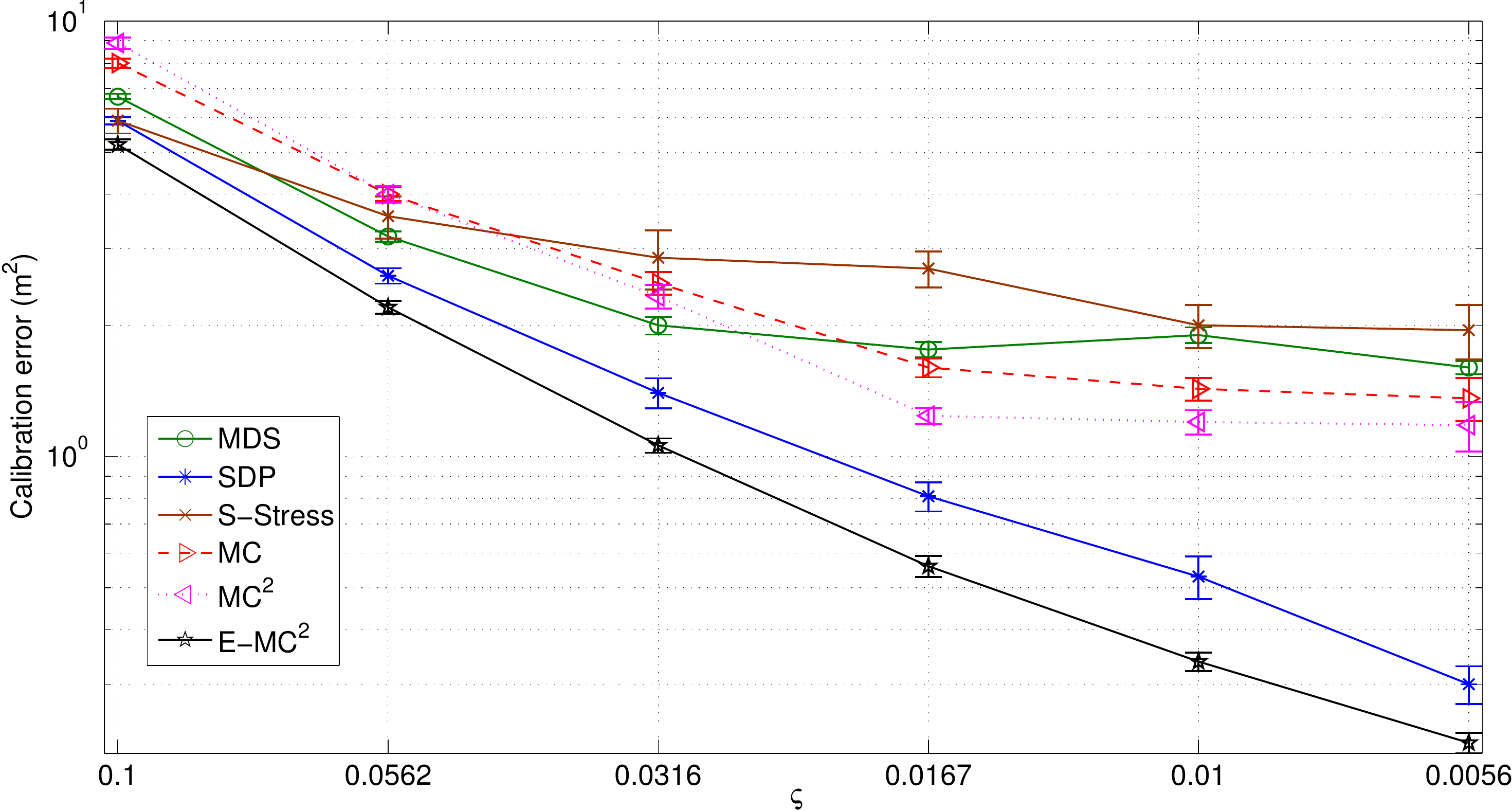} %snr-crop
\end{center}
\caption{{\color{blue}Calibration error (logarithmic scale) as quantified in~\eqref{distance} versus $\varsigma$. The error bars correspond to one standard deviation from the mean estimates.}}
\label{fig:snr_sim}
\end{figure}
%%%%%%%%%%%%%%%%%%%%%%%%%%%%%%%%%%%%%%%%%%%%%%%%%%%%%%%%%%%%%
% 
\newpage
% %%%%%%%%%%%%%%%%%%%%%%   FIGURE 6  %%%%%%%%%%%%%%%%%%%%%%%%%%
\begin{figure}[ht]
 \begin{center}
  \includegraphics[width=1\linewidth]{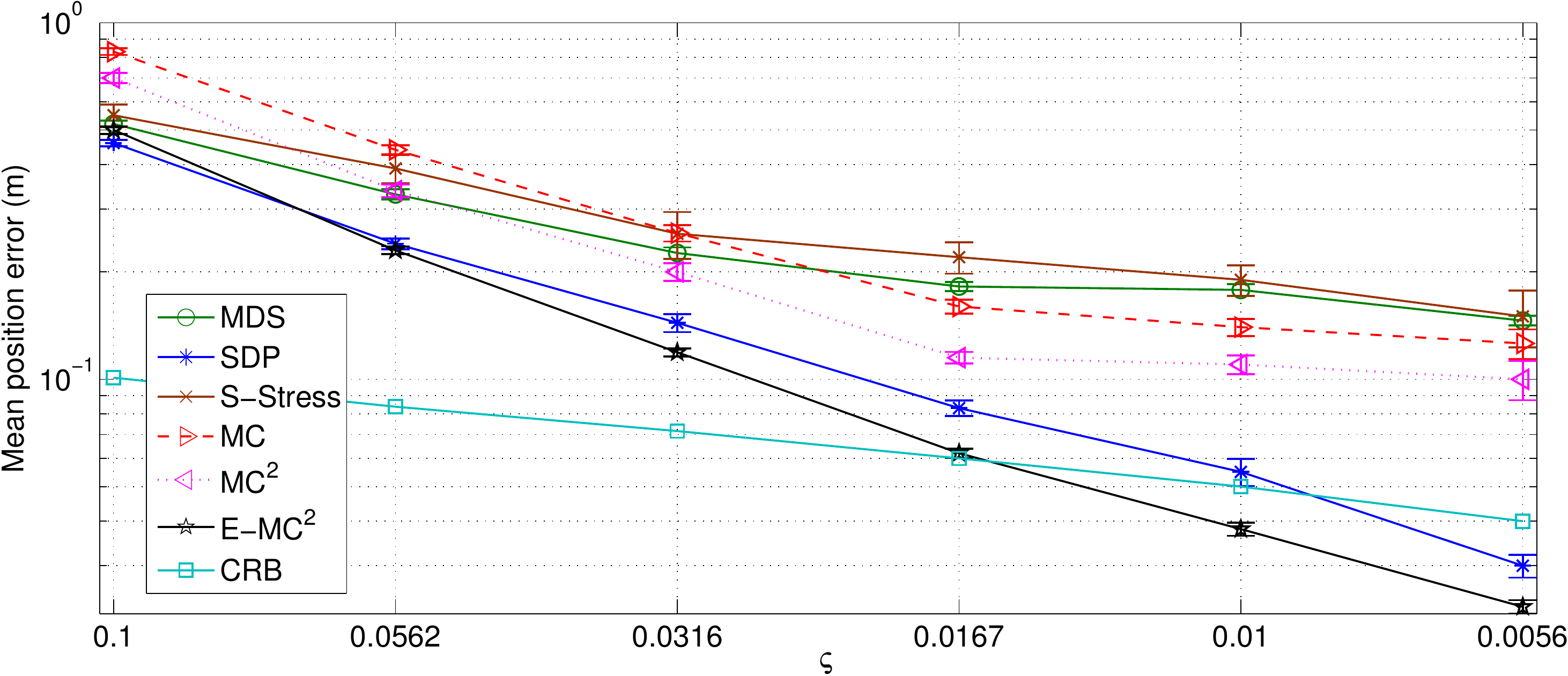}   %snr_crb-crop
\end{center}
\caption{{\color{blue}Mean position error (logarithmic scale) as defined in~\eqref{eq:pos-err} versus $\varsigma$. The error bars correspond to one standard deviation from the mean estimates.}}
\label{fig:snr_crb_sim}
\end{figure}
% %%%%%%%%%%%%%%%%%%%%%%%%%%%%%%%%%%%%%%%%%%%%%%%%%%%%%%%%%%%%%
% 
\newpage
%%%%%%%%%%%  FIGURE 7 %%%%%%%%%%%%%%%%%%%%%%%%%%%%%%%%
\begin{figure}[ht]
 \begin{center}
  \includegraphics[width=1\linewidth]{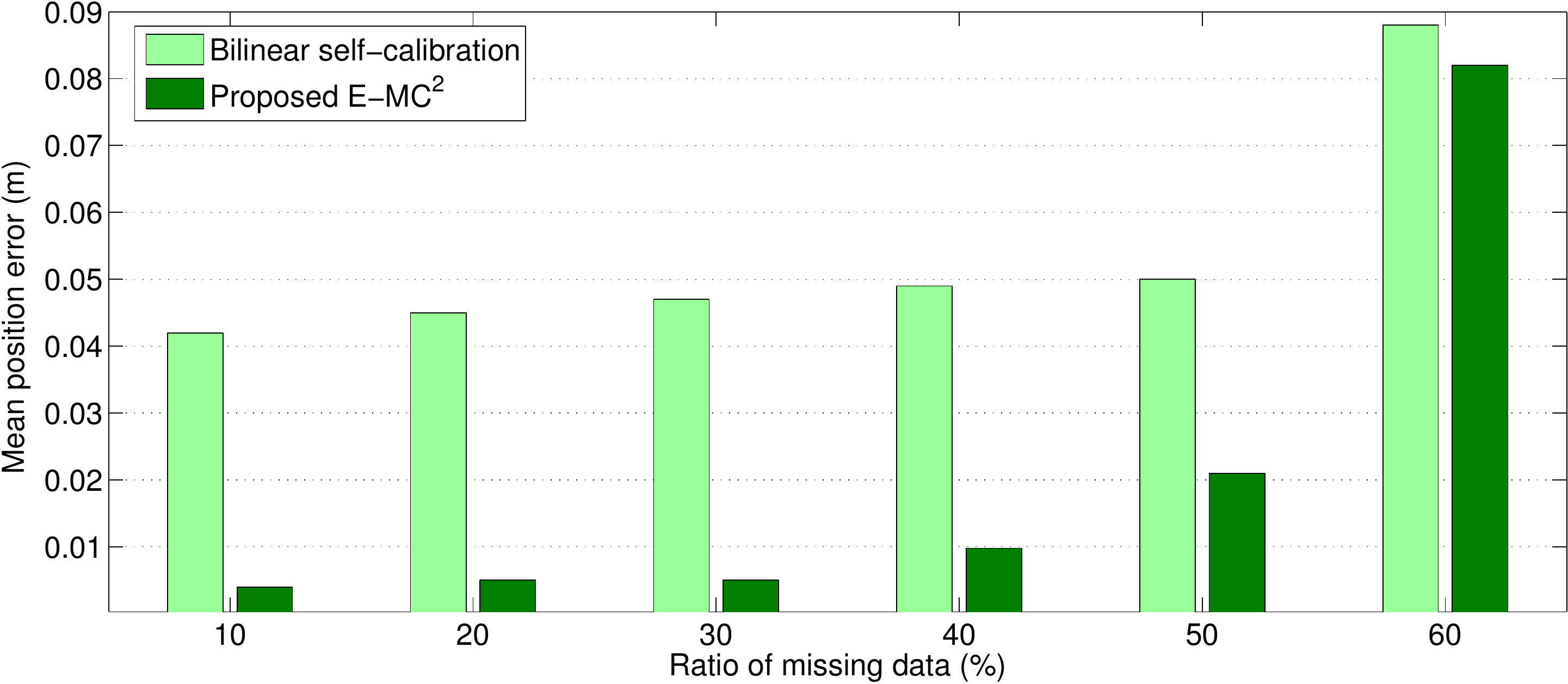}  %missing-crop
\end{center}
\caption{{\color{blue}Mean position calibration error versus the ratio of missing pairwise distances for 30 sources and 30 microphones {\color{purple}(60 nodes in total)} considered in the self-calibration method~\cite{crocco2012bilinear} and 60 microphones used for the proposed E-MC$^2$ algorithm. The standard deviation of noise in pairwise distance estimation is 0.02.}}
\label{fig:missing}
\end{figure}
%%%%%%%%%%%%%%%%%%%%%%%%%%%%%%%%%%%%%%%%%%%%%%%%%%%%%
% 
\newpage
%%%%%%%%%%       FIGURE 8  %%%%%%%%%%%%%%%%%%%%%%%%%%%%%%%%
\begin{figure}[ht]
 \begin{center}
  \includegraphics[width=1\linewidth]{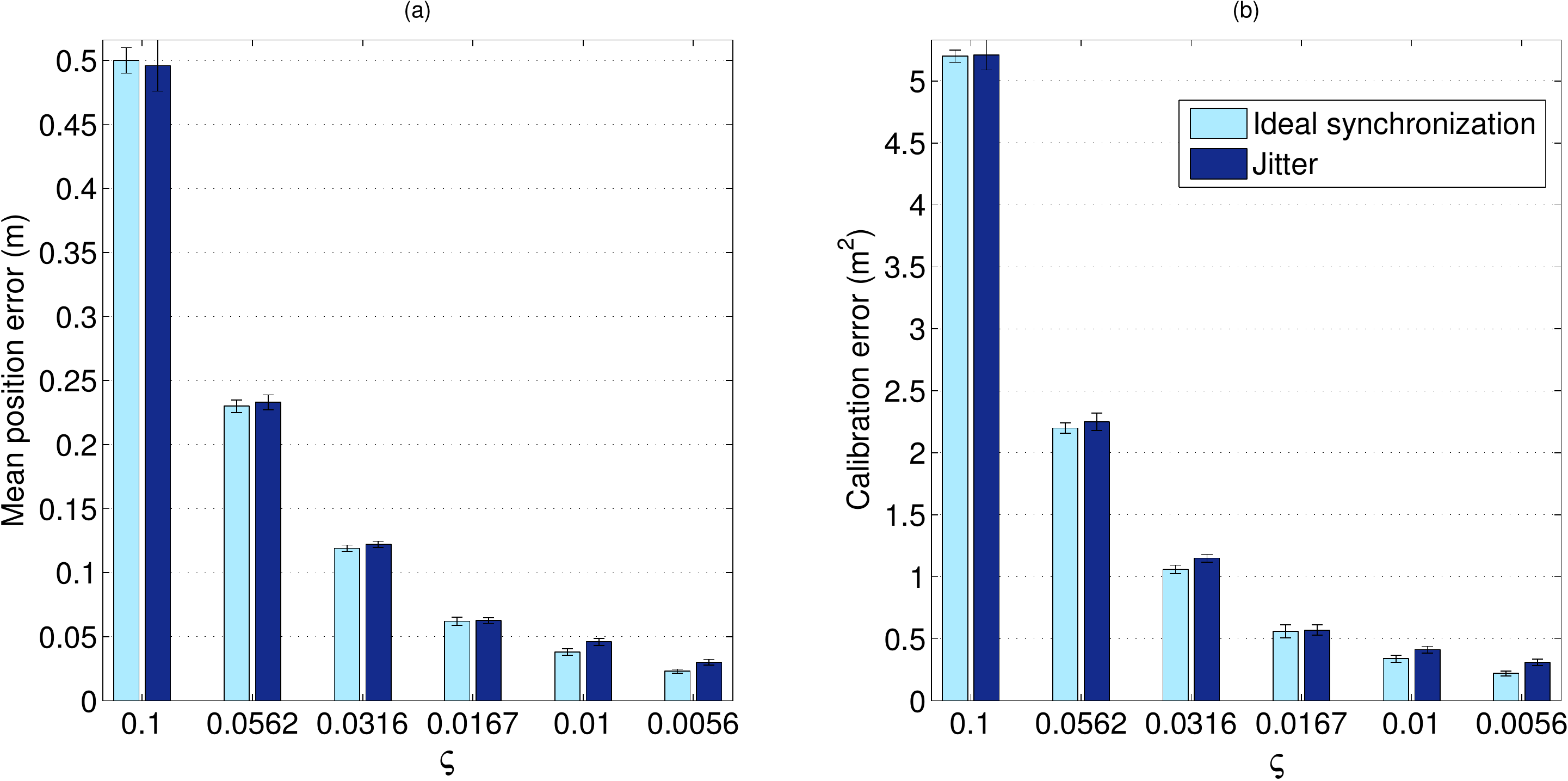}  %jitter-crop
\end{center}
\caption{{\color{blue}Effect of jitter on E-MC$^2$ algorithm quantified in terms of (a) mean position error as defined in~\eqref{eq:pos-err} as well as (b) calibration error as defined in~\eqref{distance} versus $\varsigma$. The error bars correspond to one standard deviation from the mean estimates. The number of microphones is 45 and 60\% of the pairwise distances are missing.}}
\label{fig:jitter}
\end{figure}
%%%%%%%%%%%%%%%%%%%%%%%%%%%%%%%%%%%%%%%%%%%%%%%%%%%%%%%%%%
% 
\newpage
%%%%%%%%%%%%%%%%%%      TABLE 9  %%%%%%%%%%%%%%%%%%%%%%%%%%%%%%%%%%%%%%
\begin{table}[ht]%\footnotesize
\centering
\caption{Performance of microphone array calibration in two scenarios. (1) Scenario 18-mic: two sets of 9-channel circular microphone array of diameter 20 cm; the center of both compact arrays are 1 m apart, and (2) Scenario 15-mic: a circular 9-channel microphone array of diameter 20 cm is located inside another 6-channel circular array of diameter 70 cm. The mean \emph{position} error (cm) and the \emph{calibration} error (cm$^2$) as defined in~\eqref{distance} are evaluated for different methods . The numbers in parenthesis corresponds to the error in position estimation if the experiments are repeated and averaged over 25 trials.}
\begin{tabular}{@{}c*2{c}c>{}c@{}}
\toprule[1pt]
& \multicolumn{2}{c}{\head{Scenario 18-mic}}
& \multicolumn{2}{c}{\head{Scenario 15-mic}}\\
& \normal{\head{Position (cm)}} & \normal{\head{Calibration (cm$^2$)}}
& \normal{\head{Position (cm)}} & \head{Calibration (cm$^2$)}\\
  \cmidrule(lr){2-3}\cmidrule(l){4-5}
  \multirow{1}{*}{MDS-MAP}  & 3.3 (0.72) & 175.8 & 3.18 (0.73) & 170.5 \\
  \cmidrule(lr){2-3}\cmidrule(lr){4-5}
  \multirow{1}{*}{SDP}      & 2.1 (0.3)  & 96.3 & 4.64 (0.65) & 258.8 \\
  \cmidrule(lr){2-3}\cmidrule(lr){4-5}
  \multirow{1}{*}{S-Stress} & 6.8 (0.96) & 265 & 7.05 (0.92) & 281.5 \\
  \cmidrule(lr){2-3}\cmidrule(lr){4-5}
  \multirow{1}{*}{MC} & 6.9 (1.35) & 272 & 7.5 (1.55) & 305 \\
  \cmidrule(lr){2-3}\cmidrule(lr){4-5}
  \multirow{1}{*}{MC$^2$} & 6.56 (0.91) & 225.1 & 6.8 (0.94) & 274 \\
  \cmidrule(lr){2-3}\cmidrule(lr){4-5}
  \multirow{1}{*}{E-MC$^2$} & 1.58 (0.37) & 95.5 & 1.71 (0.41) & 105.83 \\
\bottomrule[1pt]
\end{tabular}\label{tab:distributed}
\end{table}
%%%%%%%%%%%%%%%%%%%%%%%%%%%%%%%%%%%%%%%%%%%%%%%%%%%%%%%%%%%%%%%%%%%%%%%%%%%%%%%%%%
% 
\newpage
%%%%%%%%%%%%%   FIGURE 10 %%%%%%%%%%%%%%%%%%%%%%%%%%%%
\begin{figure}[ht]
 \begin{center}
  \includegraphics[width=1\linewidth]{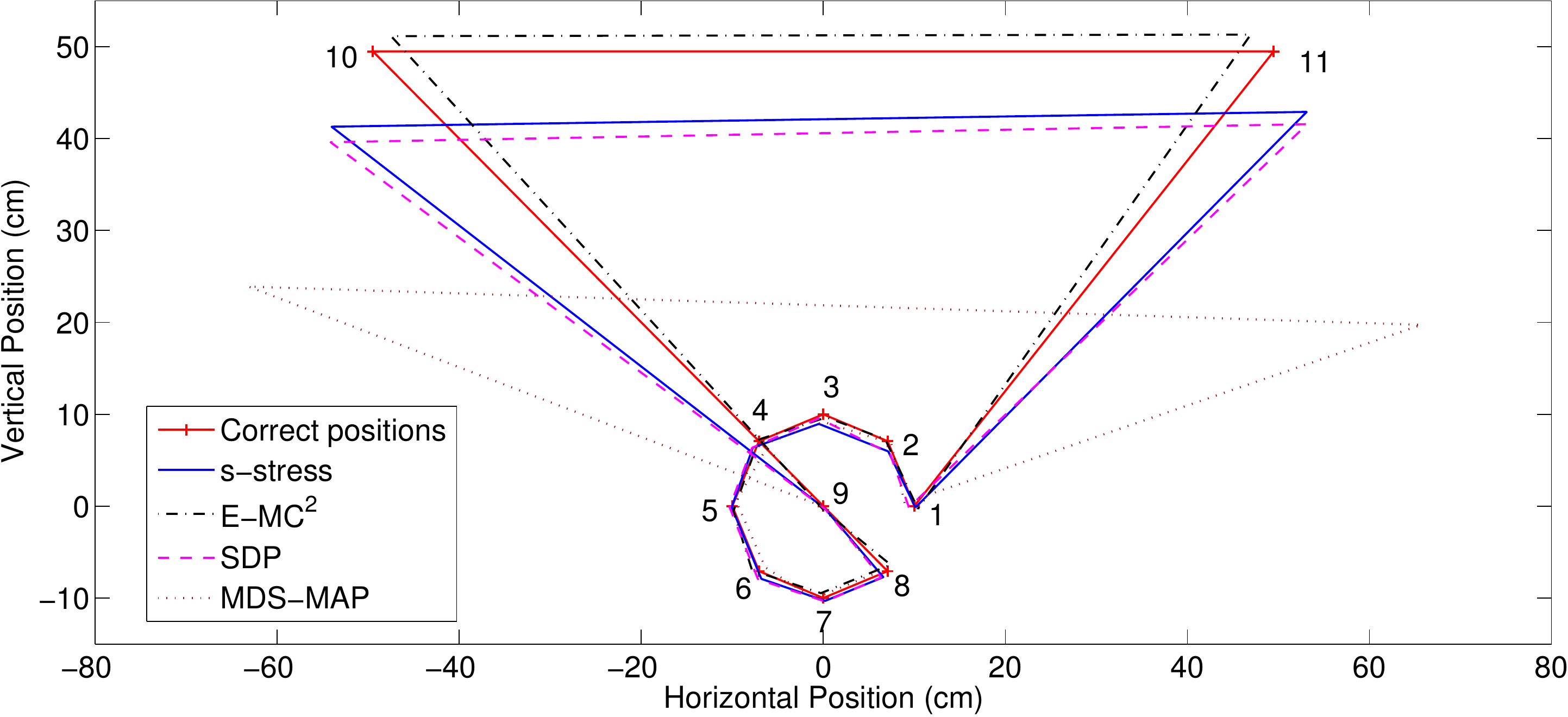}  % 2fig11mic_jour2_old-crop
\end{center}
\caption{Calibration of the eleven-element microphone array while several pairwise distances are missing.~The geometries are estimated using MDS-MAP, SDP, S-stress and the proposed proposed algorithm E-MC$^2$.}
\label{fig:11mic_old}
\end{figure}
%%%%%%%%%%%%%%%%%%%%%%%%%%%%%%%%%%%%%%%%%%%%%%%%%%%%%%%
% 
\newpage
% %%%%%%%%%%%%%   FIGURE 11 %%%%%%%%%%%%%%%%%%%%%%%%%%%%
\begin{figure}[ht]
\centering
\includegraphics[width=1\linewidth]{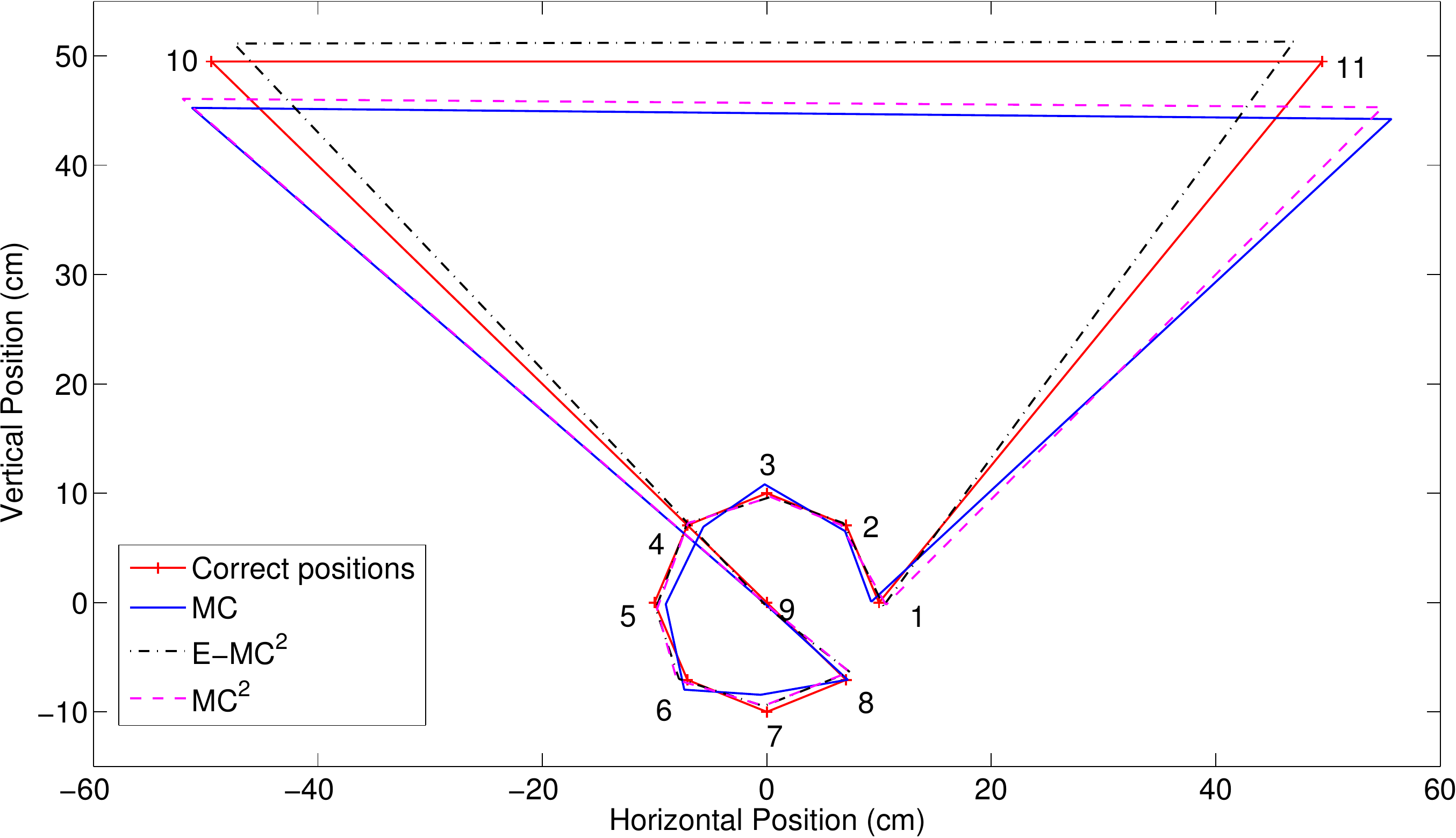}  % 4fig11mic_jour2_adv-crop.pdf
\caption{Calibration of the eleven-element microphone array while several pairwise distances are missing. The geometries are estimated using MC, MC+Cadzow (MC$^2$), and the proposed algorithm E-MC$^2$.}
\label{fig:11mic_adv}
\end{figure}
%%%%%%%%%%%%%%%%%%%%%%%%%%%%%%%%%%%%%%%%%%%%%%%%%%%%%%%%%
% 
\newpage
%%%%%%%%%%%%%%%%   FIGURE  12   %%%%%%%%%%%%%%%%%%%%%%%%
\begin{figure}[ht]
 \begin{center}
  \includegraphics[width=1\linewidth]{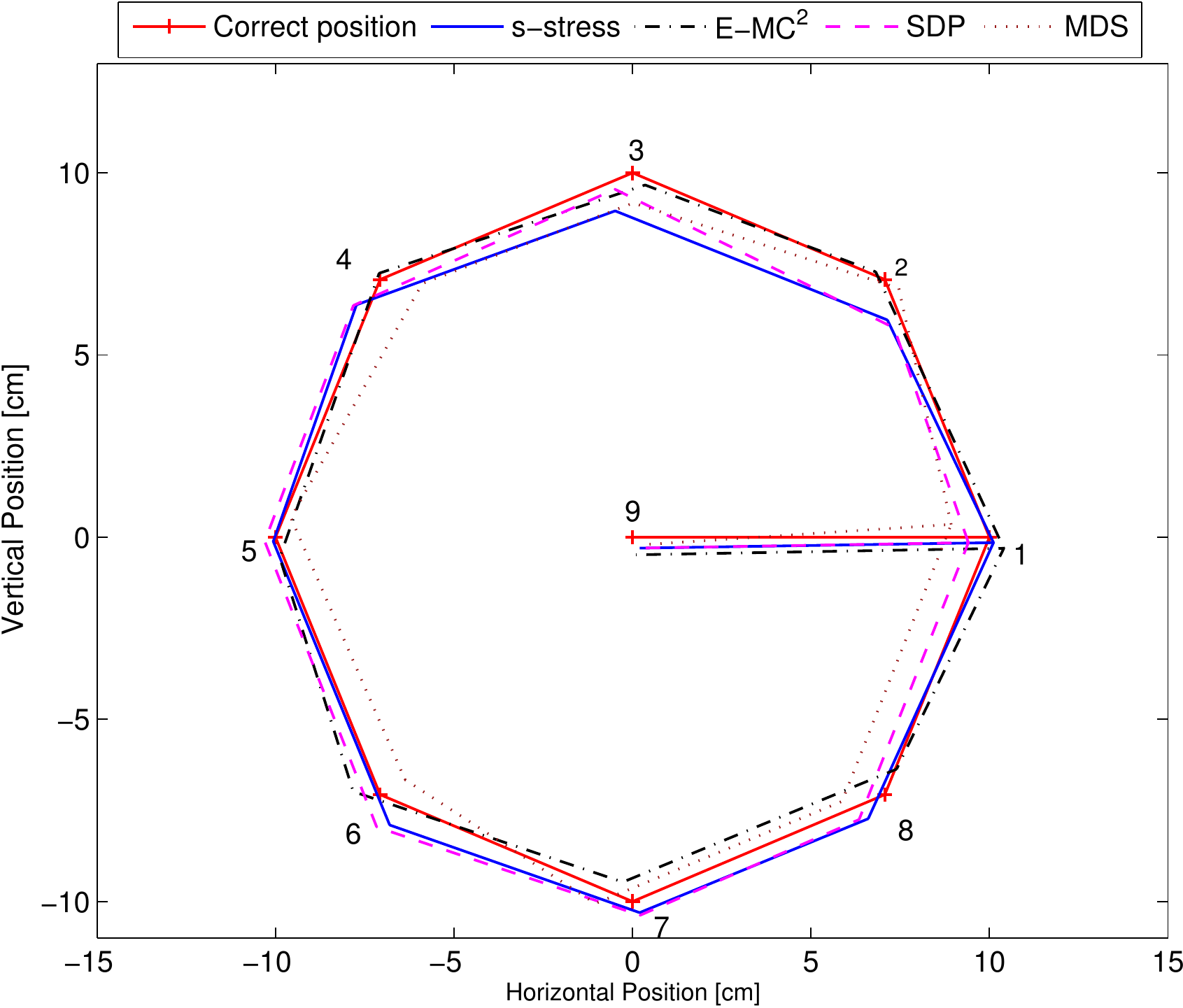}  % 3fig9mic_jour2_old-crop
 \end{center}
\caption{Calibration of the nine-element microphone array. The geometries are estimated using MDS-MAP, S-stress, SDP and the proposed Euclidean distance matrix completion algorithm, E-MC$^2$.}
\label{fig:9mic_old}
\end{figure}
%%%%%%%%%%%%%%%%%%%%%%%%%%%%%%%%%%%%%%%%%%%%%%%%%%%%%%%%
% 
\newpage
% %%%%%%%%%%%%%%%%   FIGURE  13   %%%%%%%%%%%%%%%%%%%%%%%%
\begin{figure}[ht]
 \centering
  \includegraphics[width=1\linewidth]{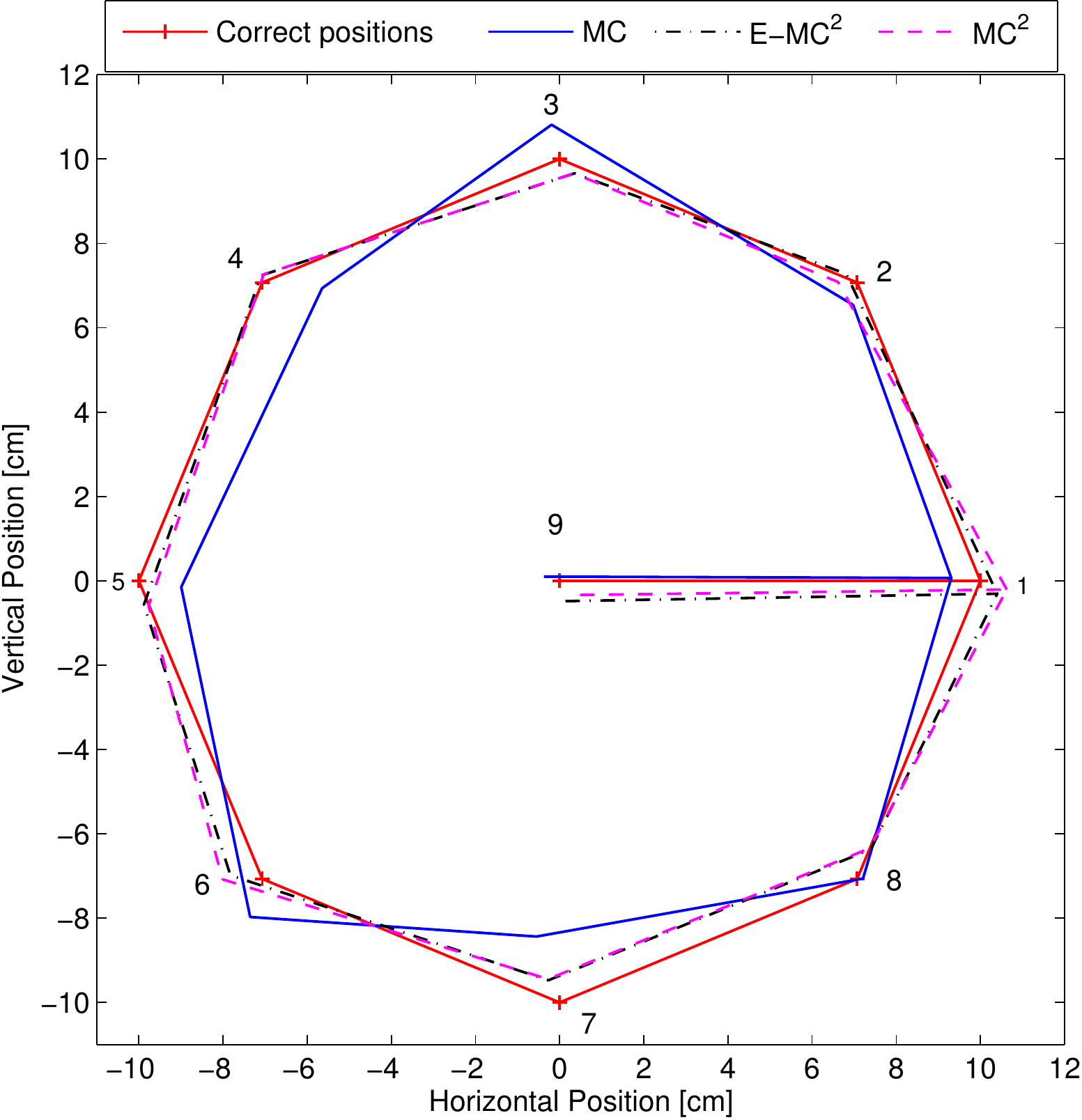} % 1fig9mic_jour2_adv-crop
\caption{{Calibration of the nine-element microphone array. The geometries are estimated using MC, MC+Cadzow (MC$^2$) and the proposed algorithm E-MC$^2$.}}
\label{fig:9mic_adv}
\end{figure}
% %%%%%%%%%%%%%%%%%%%%%%%%%%%%%%%%%%%%%%%%%%%%%%%%%%%%%%%%
% % 
\newpage
% %%%%%%%%%%%%%%%%   FIGURE  14   %%%%%%%%%%%%%%%%%%%%%%%%
\begin{table}[ht]%\footnotesize
\centering
\caption{{\color{blue} Calibration errors (cm$^2$) as defined in~\eqref{distance} for different methods of microphone array calibration.}}
\begin{tabular}{@{}c*2{c}
  c>{}c@{}}
\toprule[1pt]
& \multicolumn{2}{c}{\head{Known}}
& \multicolumn{2}{c}{\head{Missing}}\\
& \normal{\head{8-mic}} & \normal{\head{9-mic}}
& \normal{\head{11-mic}} & \head{12-mic}\\
  \cmidrule(lr){2-3}\cmidrule(l){4-5}
  \multirow{1}{*}{MDS-MAP} & 9 & 8.13 & 434.4 & 472 \\
  \cmidrule(lr){2-3}\cmidrule(lr){4-5}
  \multirow{1}{*}{SDP} & 9.09 & 8.63 & 141 & 135 \\
  \cmidrule(lr){2-3}\cmidrule(lr){4-5}
  \multirow{1}{*}{S-Stress} & 6.86 & 6.14 & 125 & 95 \\
  \cmidrule(lr){2-3}\cmidrule(lr){4-5}
  \multirow{1}{*}{MC} & 10.6 & 9.75 & 133 & 115 \\
  \cmidrule(lr){2-3}\cmidrule(lr){4-5}
  \multirow{1}{*}{MC$^2$} & 9.2 & 7.68 & 119 & 52 \\
  \cmidrule(lr){2-3}\cmidrule(lr){4-5}
  \multirow{1}{*}{E-MC$^2$} & 6.5 & 5.85 & 49.6 & 46 \\
\bottomrule[1pt]
\end{tabular}\label{tab:all-theory-cal}
\end{table}
%%%%%%%%%%%%%%%%%%%%%%%%%%%%%%%%%%%%%%%%%%%%%%%%%%%%%%%%%%%%
% 
\newpage
% %%%%%%%%%%%%%%%%   FIGURE  15   %%%%%%%%%%%%%%%%%%%%%%%%
\begin{table}[ht]%\footnotesize
\centering
\caption{{\color{blue} Position estimation errors (cm) as defined in~\eqref{eq:pos-err} for different methods of microphone array calibration.}}
\begin{tabular}{@{}c*2{c}
  c>{}c@{}}
\toprule[1pt]
& \multicolumn{2}{c}{\head{Known}}
& \multicolumn{2}{c}{\head{Missing}}\\
& \normal{\head{8-mic}} & \normal{\head{9-mic}}
& \normal{\head{11-mic}} & \head{12-mic}\\
  \cmidrule(lr){2-3}\cmidrule(l){4-5}
  \multirow{1}{*}{MDS-MAP} & 0.83 & 0.78 & 6.34 & 7.23 \\
  \cmidrule(lr){2-3}\cmidrule(lr){4-5}
  \multirow{1}{*}{SDP} & 0.86 & 0.81 & 2.88 & 2.35 \\
  \cmidrule(lr){2-3}\cmidrule(lr){4-5}
  \multirow{1}{*}{S-Stress} & 0.69 & 0.61 & 2.5 & 1.9 \\
  \cmidrule(lr){2-3}\cmidrule(lr){4-5}
  \multirow{1}{*}{MC} & 1.1 & 0.97 & 2.6 & 2.1 \\
  \cmidrule(lr){2-3}\cmidrule(lr){4-5}
  \multirow{1}{*}{MC$^2$} & 0.91 & 0.74 & 2.16 & 1.7 \\
  \cmidrule(lr){2-3}\cmidrule(lr){4-5}
  \multirow{1}{*}{E-MC$^2$} & 0.64 & 0.97 & 1.06 & 1 \\
\bottomrule[1pt]
\end{tabular}\label{tab:all-theory-pos}
\end{table}
%%%%%%%%%%%%%%%%%%%%%%%%%%%%%%%%%%%%%%%%%%%%%%%%%%%%%%%%%
% 
\newpage
%%%%%%%%%  FIGURE LAST    %%%%%%%%%%%%%%%%%%%%
\begin{figure}[ht]
 \begin{center}
  \includegraphics[width=1\linewidth]{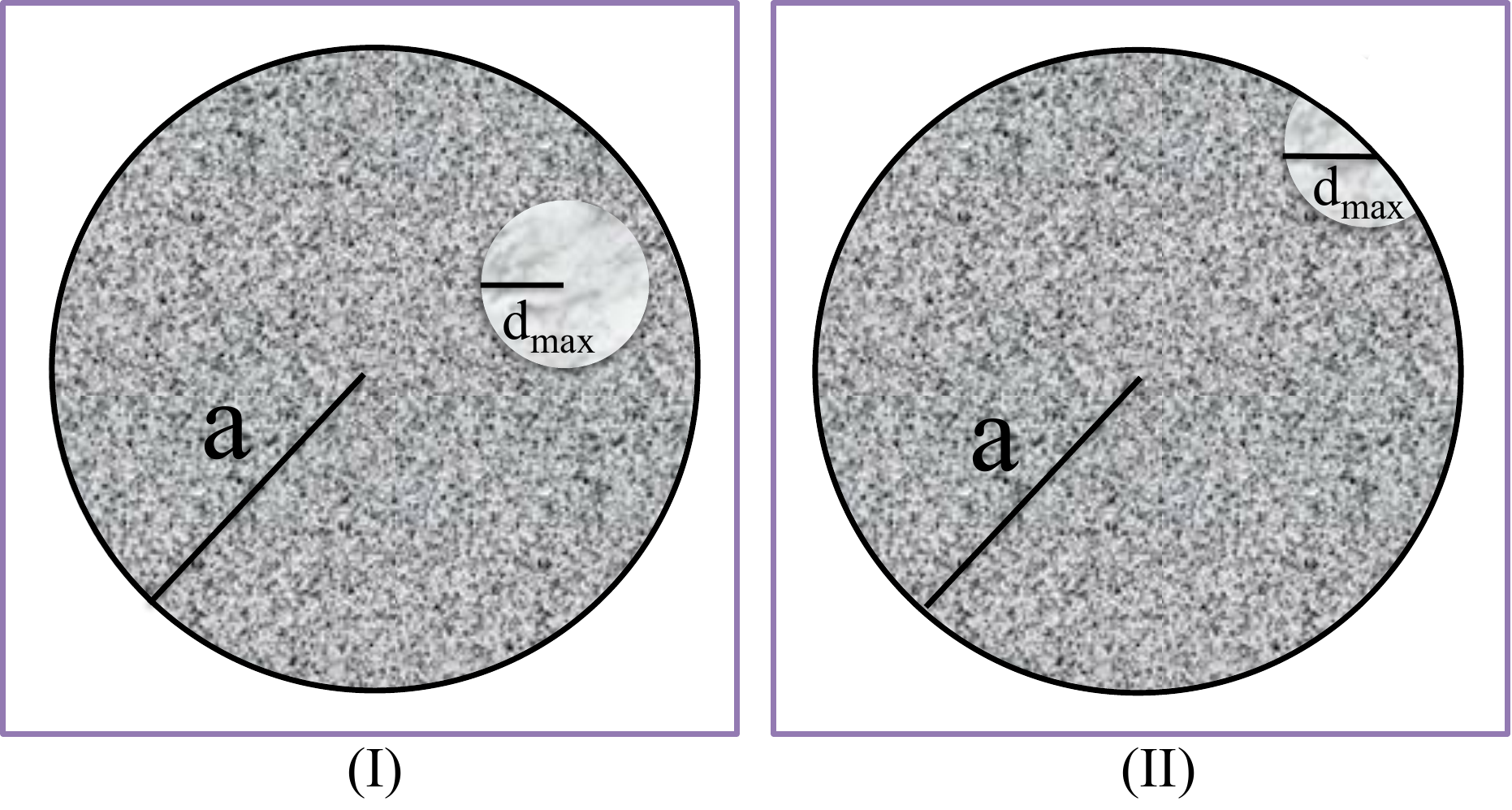}   %Amin-Amax
 \end{center}
\caption{Scenario corresponding to the (I) lower bound and (II) upper bound of the probability $q$ of structured missing distances.}
\label{fig:bounding}
\end{figure}
%%%%%%%%%%%%%%%%%%%%%%%%%%%%%%%%%%%%%%%%%%%%%%%

\end{document}